\documentclass[a4paper]{jpconf}
\usepackage{graphicx}
\begin{document}
\title{Testing the AGN Unification Model in the Infrared}

\author{C. Ramos Almeida$^{1,2}$, N. A. Levenson$^3$, A. Alonso-Herrero$^{4,5}$, A. Asensio Ramos$^{1,2}$, J. M. Rodr\'\i guez Espinosa$^{1,2}$, 
A. M. P\'erez Garc\' ia$^{1,2}$, C. Packham$^6$, R. Mason$^7$, J. T. Radomski$^3$, and T. D\' iaz-Santos$^8$}

\address{
$^1$ Instituto de Astrof\'\i sica de Canarias, La Laguna, Tenerife, E-38205, Spain.  \\
$^2$ Departamento de Astrof\' isica, Universidad de La Laguna, E-38205, Spain. \\
$^3$ Gemini Observatory, Casilla 603, La Serena, Chile.                        \\ 
$^4$ Instituto de F\'isica de Cantabria, CSIC-UC, Santander, E-39005, Spain.   \\
$^5$ Augusto Gonz\'alez Linares Senior Research Fellow.			       \\	
$^6$ Astronomy Department, University of Florida, Gainesville, FL 32611-2055, USA. \\
$^7$ Gemini Observatory, Northern Operations Center, Hilo, HI 96720, USA.       \\
$^8$ Spitzer Science Center, Pasadena, CA 91125, USA.}

\ead{cra@iac.es}

\begin{abstract}
We present near-to-mid-infrared spectral energy distributions (SEDs) for 21 Seyfert galaxies,
using subarcsecond resolution imaging data.  
Our aim is to compare the properties Seyfert 1 (Sy1) and Seyfert 2 (Sy2) tori using clumpy torus models 
and a Bayesian approach to fit the infrared (IR) nuclear SEDs. 
These dusty tori have physical sizes smaller than 6 pc radius, as derived from our fits. 
Active galactic nuclei (AGN) unification schemes account for a variety of observational differences in terms of 
viewing geometry. However, we find evidence that strong unification may not hold,
and that the immediate dusty surroundings of Sy1 and Sy2 nuclei are intrinsically 
different. The Type 2 tori studied here are broader, have more clumps, and these clumps have 
lower optical depths than those of Type 1 tori. The larger the covering factor of the torus, 
the smaller the probability of having direct view of the AGN, and vice-versa. In our sample, 
Sy2 tori have larger covering factors (C$_T$=0.95$\pm$0.02) and smaller escape probabilities 
than those of Sy1 (C$_T$=0.5$\pm$0.1). 
Thus, on the basis of the results presented here, the classification of a Seyfert galaxy 
may depend more on the intrinsic properties of the torus rather than on its 
mere inclination, in contradiction with the simplest unification model.

\end{abstract}

\section{Introduction}

The unified model for active galaxies \cite{Antonucci93} is based on the 
existence of a dusty toroidal structure surrounding the central region of AGN. 
This geometry of the obscuring material allows the central engines of Type 1 AGN to be seen directly, 
resulting in typical spectra with both narrow and broad emission lines, whereas 
in Type 2 AGN the broad-line region (BLR) is obscured.

Pioneering work in modelling the dusty torus \cite{Pier92,Pier93,Granato94,Efstathiou95,Granato97} 
assumed a uniform dust density distribution, although from the start, \cite{Krolik88} realized
that smooth dust distributions cannot survive within the AGN vicinity. 
In order to solve the discrepancies between the observations and the predictions from smooth torus models, 
an intensive search for an alternative torus geometry has been carried out in the last decade. 
The clumpy dusty torus models \cite{Nenkova08a,Nenkova08b,Honig06,Schartmann08}
propose that the dust is distributed in clumps, instead of homogeneously filling the torus volume, and they 
are making significant progress in accounting for the mid-IR (MIR) emission of AGNs 
\cite{Mason06,Mor09,Horst09,Nikutta09,Ramos09a,Ramos11,Honig10,Alonso11}.

In this work, we report IR nuclear fluxes for a representative 
sample of nearby Seyfert galaxies, for which we estimate unresolved nuclear MIR fluxes.
We compile near-IR (NIR) nuclear fluxes from the literature of similar resolution to construct nuclear 
SEDs for all the galaxies, and fit them with clumpy torus models, which we interpolate from the 
\cite{Nenkova08a,Nenkova08b} database.
Table 1 summarizes key observational properties of the sources in the sample. 

\begin{table}[h]
\caption{\label{sources}Basic Galaxy Data.}
\begin{center}
\begin{footnotesize}
\begin{tabular}{lcccccccc}
\br
Galaxy & Seyfert & $z$ & Distance & Scale & Telescope & Instrument & Filters & Resolution \\
 & type & & (Mpc) & (pc~arcsec$^{-1}$) & & & & (arcsec) \\ 
\mr
NGC 1097    &  1	  &  0.0042 &	19   & 92     & Gemini-S & T-ReCS  & Si-5, Qa  & 0.41, 0.52 \\  
NGC 1566    &  1	  &  0.0050 &	20   & 97     & Gemini-S & T-ReCS  & Si-2, Qa  & 0.30, 0.53 \\  
NGC 6221    &  1	  &  0.0050 &	18   & 87     & Gemini-S & T-ReCS  & Si-2, Qa  & 0.32, 0.55 \\ 
NGC 7469    &  1	  &  0.0163 &	65   & 315    & Gemini-S & T-ReCS  & Si-2, Qa  & 0.31, 0.55 \\
\mr
NGC 3227    &  1.5	  & 0.0039  &	17   & 82     & Gemini-N & Michelle& N'        & 0.39       \\
NGC 4151    &  1.5	  & 0.0033  &	13   & 64     & Gemini-N & OSCIR   & N, IHW18  & 0.53, 0.58 \\
NGC 6814    &  1.5	  &  0.0052 &	21   & 102    & Gemini-S & T-ReCS  & Si-2, Qa  & 0.28, 0.53 \\ 
\mr
NGC 1365    &  1.8	  & 0.0055  &	18   & 107    & CTIO 4m  & OSCIR   & N, IHW18  & 0.92, 1.03 \\  
NGC 2992    &  1.9	  & 0.0077  &	31   & 149    & Gemini-N & Michelle& N', Qa    & 0.32, 0.53 \\  
NGC 5506    &  1.9	  & 0.0062  &	25   & 120    & Gemini-N & Michelle& N', Qa    & 0.36, 0.51 \\  
\mr
Centaurus A &  2	  & 0.0018 &   3.5   & 17     & Gemini-S & T-ReCS  & Si-2, Qa  & 0.30, 0.53 \\
Circinus    &  2	  & 0.0014 &   4     & 20     & Gemini-S & T-ReCS  & Si-2, Qa  & 0.33, 0.55 \\
IC 5063	    &  2	  & 0.0113 &   45    & 219    & Gemini-S & T-ReCS  & Si-2, Qa  & 0.40, 0.62 \\
Mrk 573	    &  2	  & 0.0172 &   69    & 334    & Gemini-S & T-ReCS  & N, Qa     & 0.36, 0.54 \\
NGC 1386    &  2	  & 0.0029 &   11    & 56     & Gemini-S & T-ReCS  & N, Qa     & 0.31, 0.54 \\  
NGC 1808    &  2	  & 0.0033 &   11    & 64     & CTIO 4m  & OSCIR   & N, IHW18  & 0.94, 1.02 \\
NGC 3081    &  2	  & 0.0080 &   32    & 155    & Gemini-S & T-ReCS  & Si-2, Qa  & 0.30, 0.56 \\
NGC 3281    &  2	  & 0.0107 &   43    & 208    & Gemini-S & T-ReCS  & N, Qa     & 0.34, 0.58 \\  
NGC 4388    &  2	  & 0.0084 &   34    & 163    & Gemini-N & Michelle& N', Qa    & 0.34, 0.50 \\
NGC 7172    &  2	  & 0.0087 &   35    & 169    & Gemini-S & T-ReCS  & N         & 0.51       \\  
NGC 7582    &  2	  & 0.0053 &   21    & 103    & CTIO 4m  & OSCIR   & N, IHW18  & 0.76, 0.99 \\  
\br
\end{tabular}
\end{footnotesize}
\end{center}
\end{table}

\section{MIR Observations}
\label{observations}

Ground-based MIR high-angular resolution observations of 21 nearby active galaxies were carried out
over the past years for a variety of science drivers. We make use of this archive of  data in this paper. 
The first set of observations was obtained with the University of
Florida MIR camera/spectrometer OSCIR at the 4 m
Blanco Telescope at Cerro Tololo Inter-American Observatory (CTIO) and
at the Gemini North Telescope. 
Another set of observations was performed with the MIR
camera/spectrograph T-ReCS (Thermal-Region Camera Spectrograph) 
on the Gemini-South telescope.  
The last set was obtained with the MIR
camera/spectrograph Michelle on the Gemini North
Telescope (see Table 1). 
Observations of PSF and flux standard stars were made for each galaxy through the 
same filters. The PSF star images were employed to determine the unresolved (i.e., nuclear) 
component of each galaxy. 
For further details on the MIR observations of the whole sample we 
refer the reader to \cite{Ramos09a,Ramos11}.

\section{SED Observational Properties}
\label{sed}

Using the MIR data described in Section \ref{observations} and NIR data of similar resolution 
compiled from the literature, we constructed
subarcsecond resolution nuclear SEDs in the wavelength range from $\sim$1 to 18 $\mu$m~for the galaxies in 
Table 1. Figures \ref{sy1_fits_a} and \ref{sy1_fits_b} show the individual SEDs.

Using the highest resolution data only ($\le$0.55 arcsec) we constructed average Sy1 and Sy2 templates 
(see Figure 4 in \cite{Ramos11}) and measured the 1.265--18.3 $\mu$m~IR slope
(f$_{\nu}~\alpha~\nu^{-\alpha_{IR}}$) of the Sy1 template: $\alpha_{IR} = 1.7\pm0.3$, as well as the NIR 
($\alpha_{NIR} = 1.6\pm0.2$)
and MIR spectral indexes ($\alpha_{MIR} = 2.0\pm0.2$). A flat NIR slope
indicates an important contribution of hot dust emission (up to $\sim$1000-1200 K;
\cite{Barvainis87}) that comes from the immediate vicinity of the AGN. 
The shape of the Sy2 mean SED is very steep ($\alpha_{IR} = 3.1\pm0.9$, $\alpha_{NIR}=3.6\pm0.8$, and $\alpha_{MIR}=2.0\pm0.2$) 
compared with those of the Type 1 Seyferts. In general, Sy2 have steeper 1--10 $\mu$m~SEDs than Sy1 
\cite{Fadda98,Alonso01,Alonso03}.
On the contrary, the MIR slope is the same for the Sy1 and Sy2 templates ($\alpha_{MIR}=2.0\pm0.2$).

\section{SED Modelling}
\label{modelling}

The clumpy dusty torus models of Nenkova et al.~(2002) hold that the dust
surrounding the central engine of an AGN is distributed in clumps, instead of homogeneously filling the torus volume.  
The clumpy database now contains 1.2$\times$10$^6$ models, calculated for a fine grid of model
parameters. The inherent degeneracy between these parameters has to be taken into account
when fitting the observables. To this end, we developed a Bayesian inference tool (BayesClumpy; \cite{Asensio09}).

The priors for the model parameters are assumed to be truncated uniform
distributions in the intervals reported in Table 2. Therefore, we give the same weight
to all the values in each interval. Apart from
the six parameters that characterize the models, there is an additional parameter
that accounts for the vertical displacement required to match the fluxes of a chosen model to an observed 
SED. This vertical shift scales with the AGN bolometric luminosity (see Section \ref{discussion}). 

\begin{table}[h]
\caption{Clumpy Model Parameters and Considered Intervals}
\begin{center}
\begin{small}
\begin{tabular}{lcl}
\br
Parameter & Abbreviation & Interval \\
\mr
Width of the angular distribution of clouds            & $\sigma$        & [15$^o$, 70$^o$]  \\
Radial extent of the torus                             & $Y$             & [5, 30]        \\
Number of clouds along the radial equatorial direction & $N_0$           & [1, 15] \\
Power-law index of the radial density profile          & $q$             & [0, 3]    \\
Inclination angle of the torus                         & $i$             & [0$^o$, 90$^o$]   \\
Optical depth per single cloud                         & $\tau_{V}$      & [5, 150] \\
\br         
\end{tabular}
\end{small}
\end{center}
\end{table}

The results of the fitting process of the IR SEDs with the interpolated version of the
clumpy models of \cite{Nenkova08a,Nenkova08b} are the posterior distributions for the six free 
parameters that describe the models and the vertical shift. 
When the observed data introduce sufficient information into the fit, 
the resulting posteriors will clearly differ from the input uniform priors, 
either showing trends or being centered at certain values within the intervals considered.

\begin{figure*}[!ht]
\centering
{\par
\includegraphics[width=5.2cm]{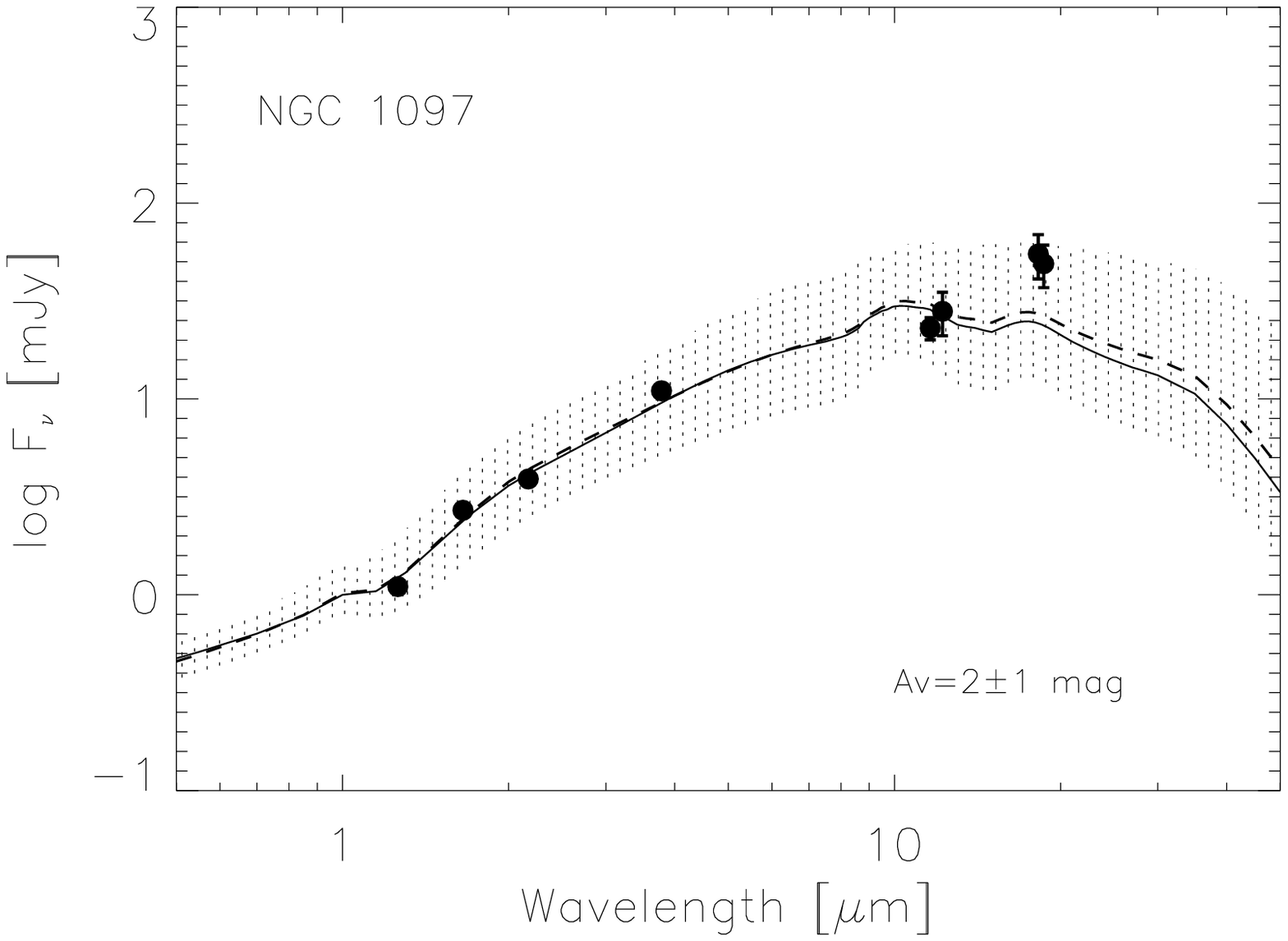}
\includegraphics[width=5.2cm]{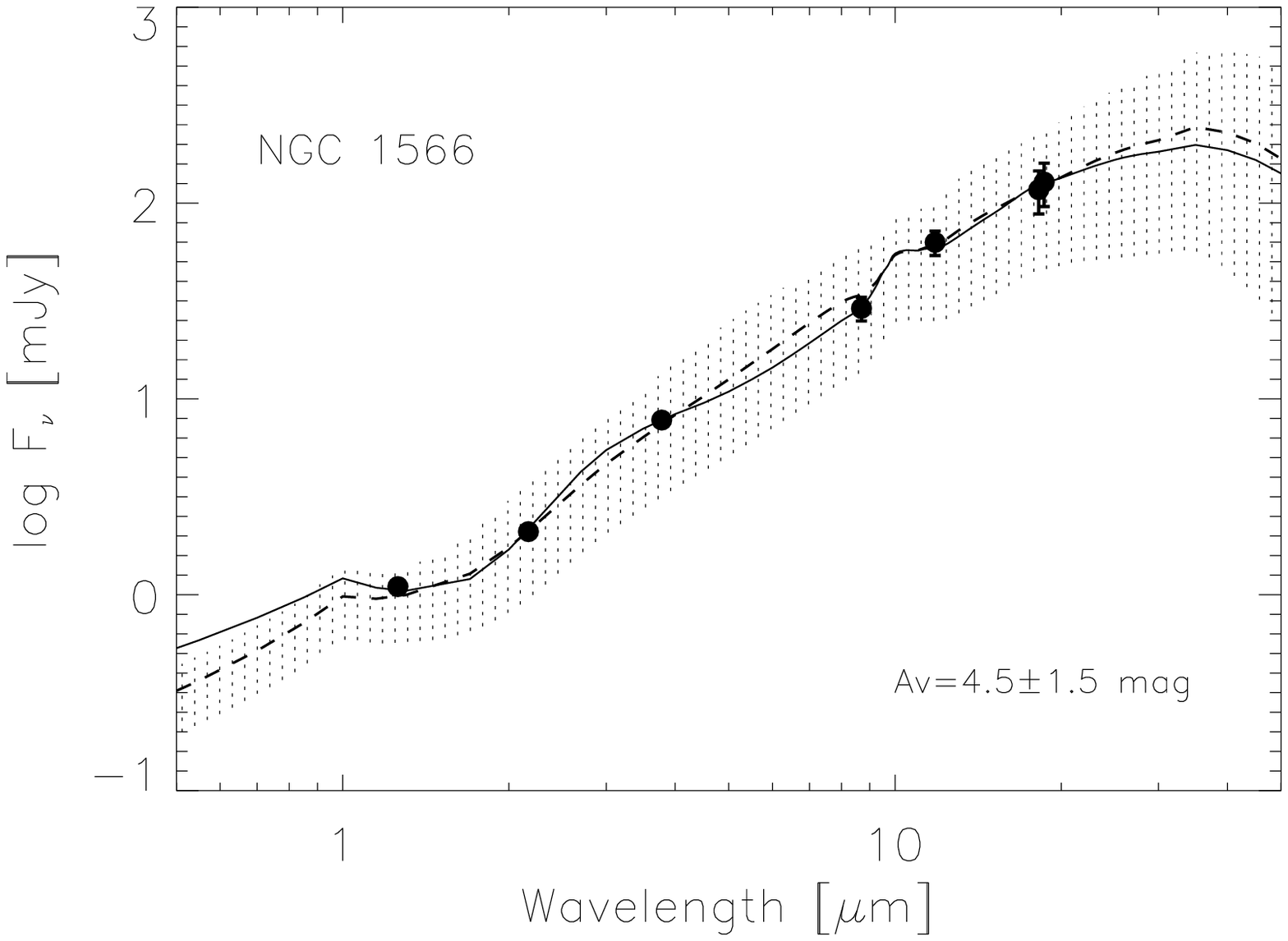}
\includegraphics[width=5.2cm]{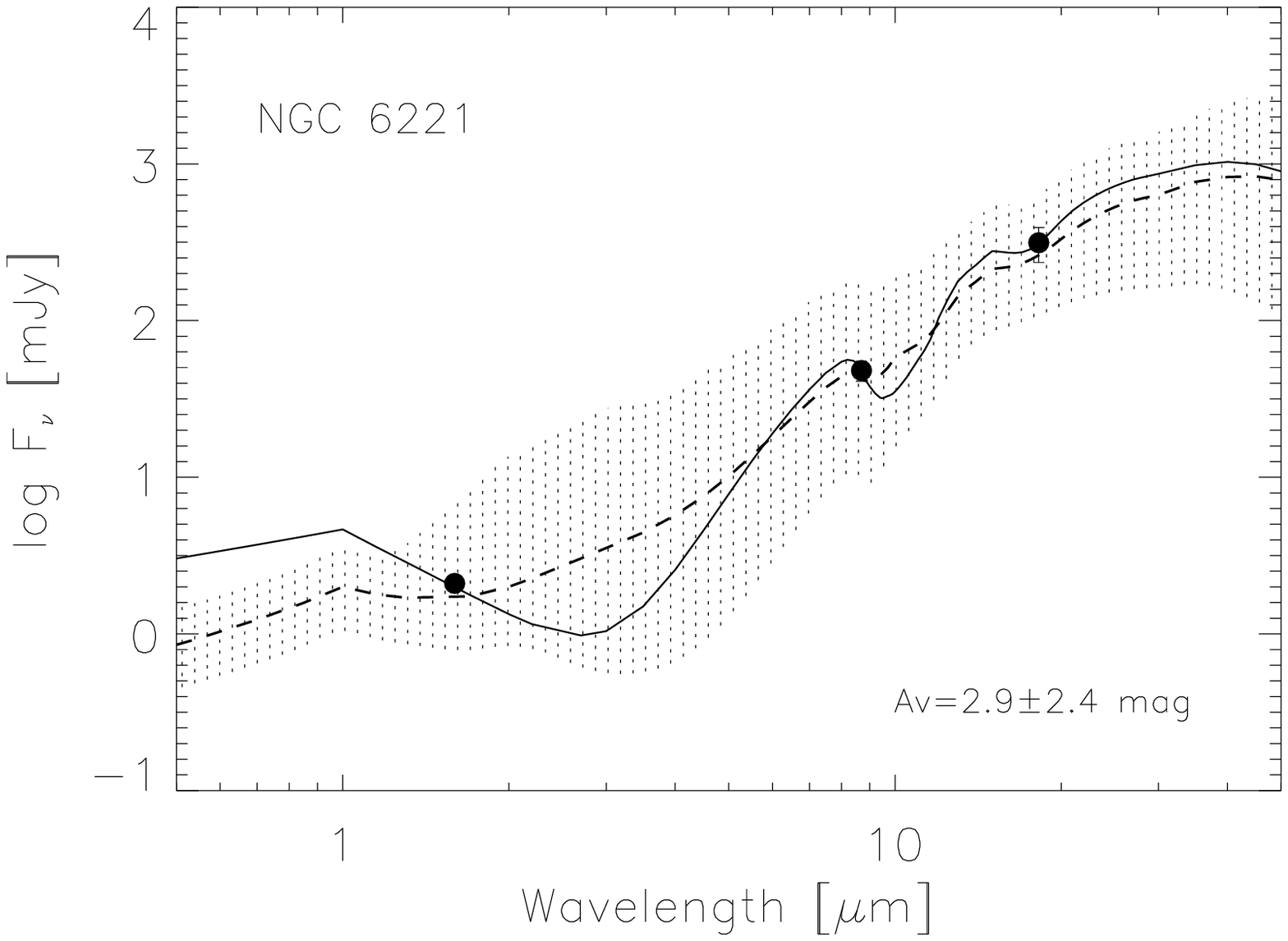}
\includegraphics[width=5.2cm]{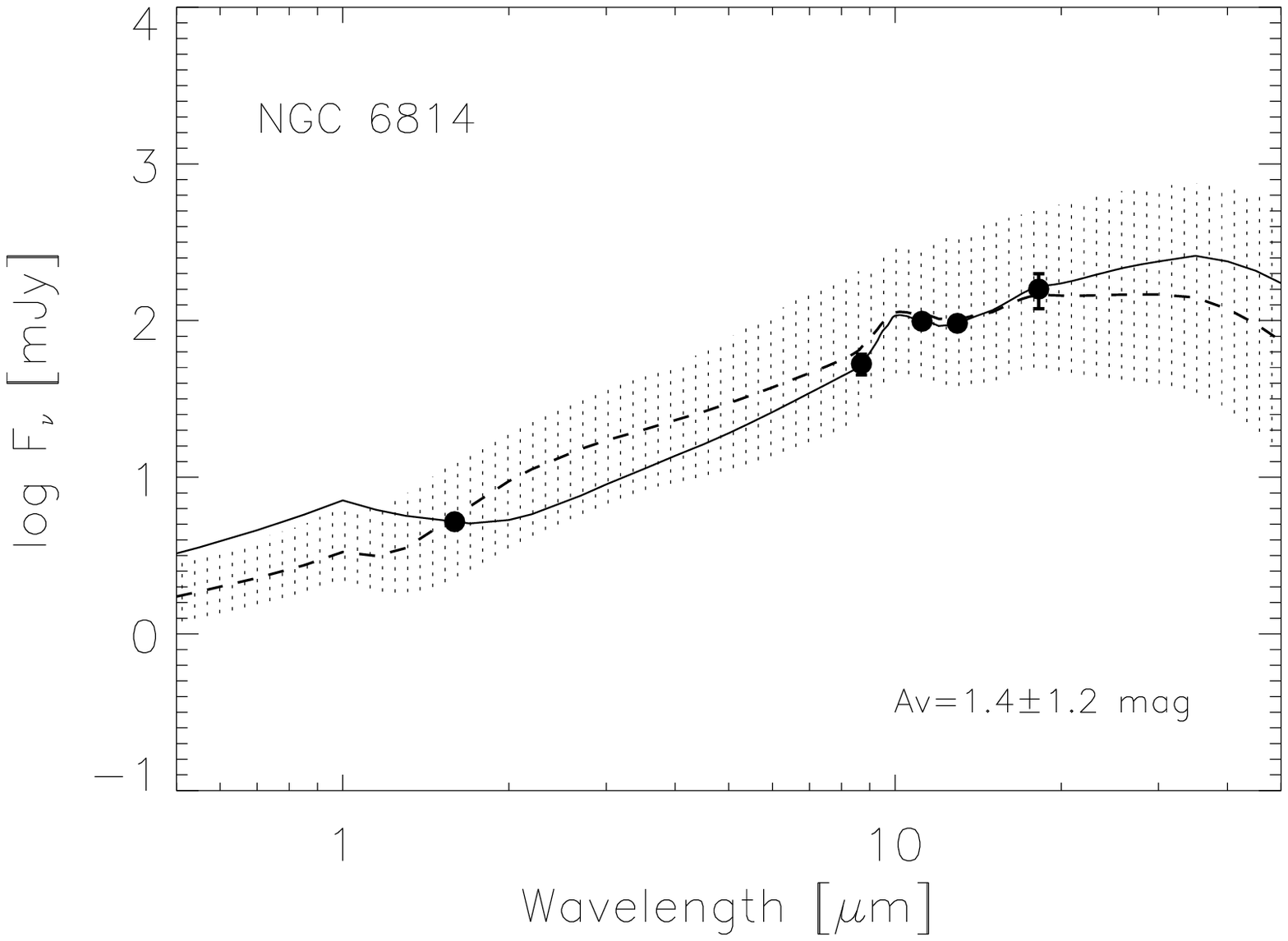}
\includegraphics[width=5.2cm]{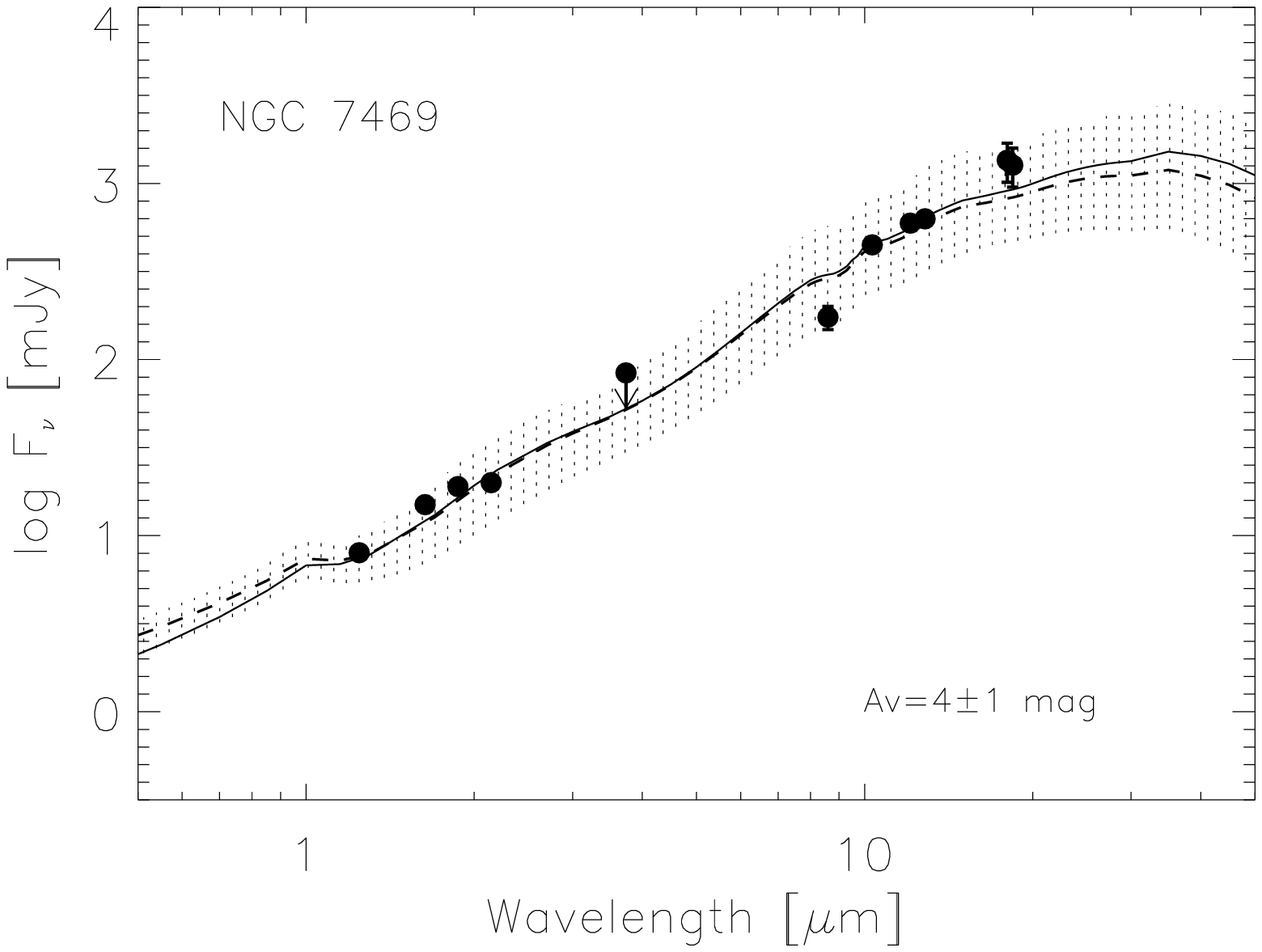}
\includegraphics[width=5.2cm]{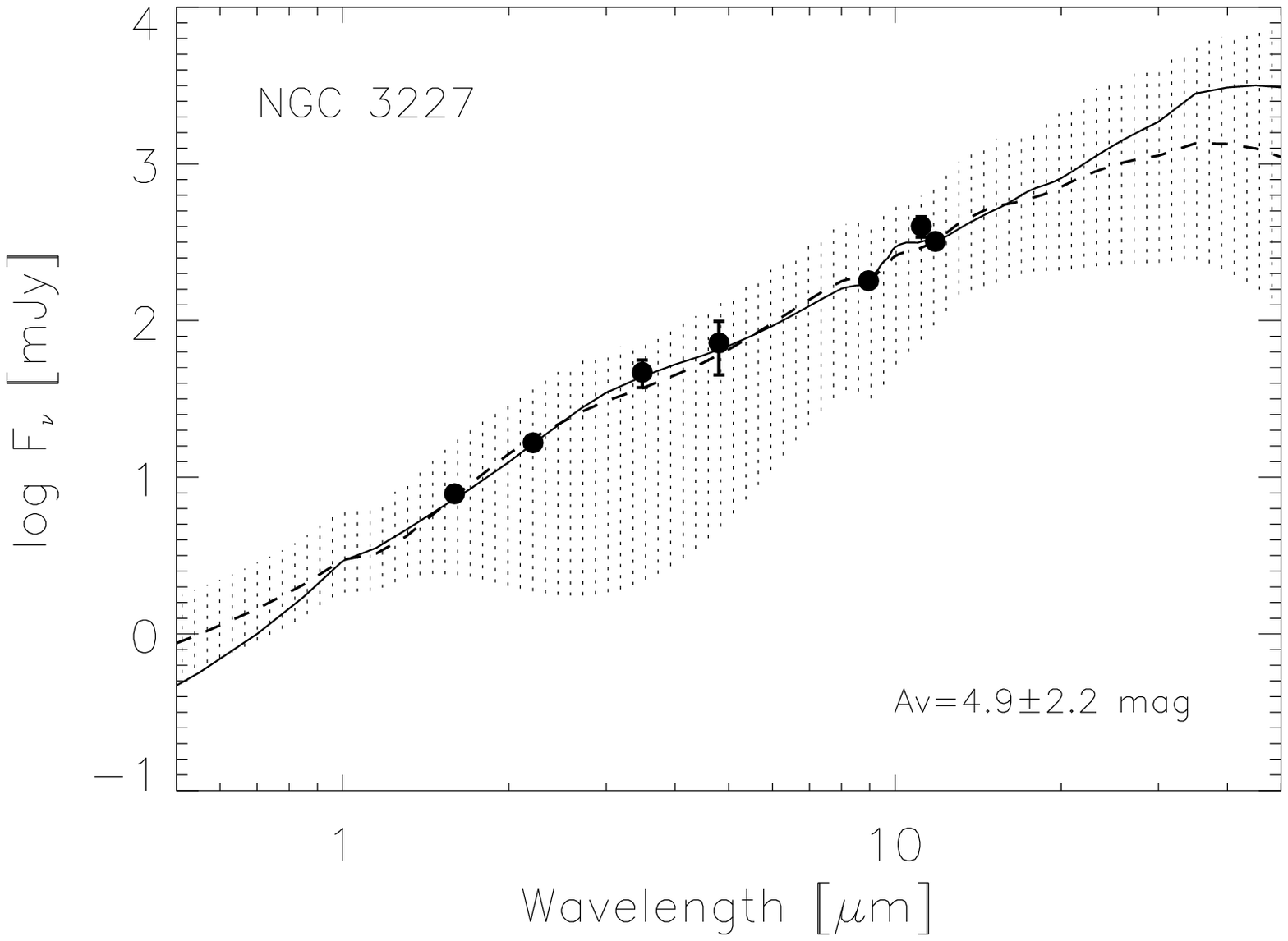}
\includegraphics[width=5.2cm]{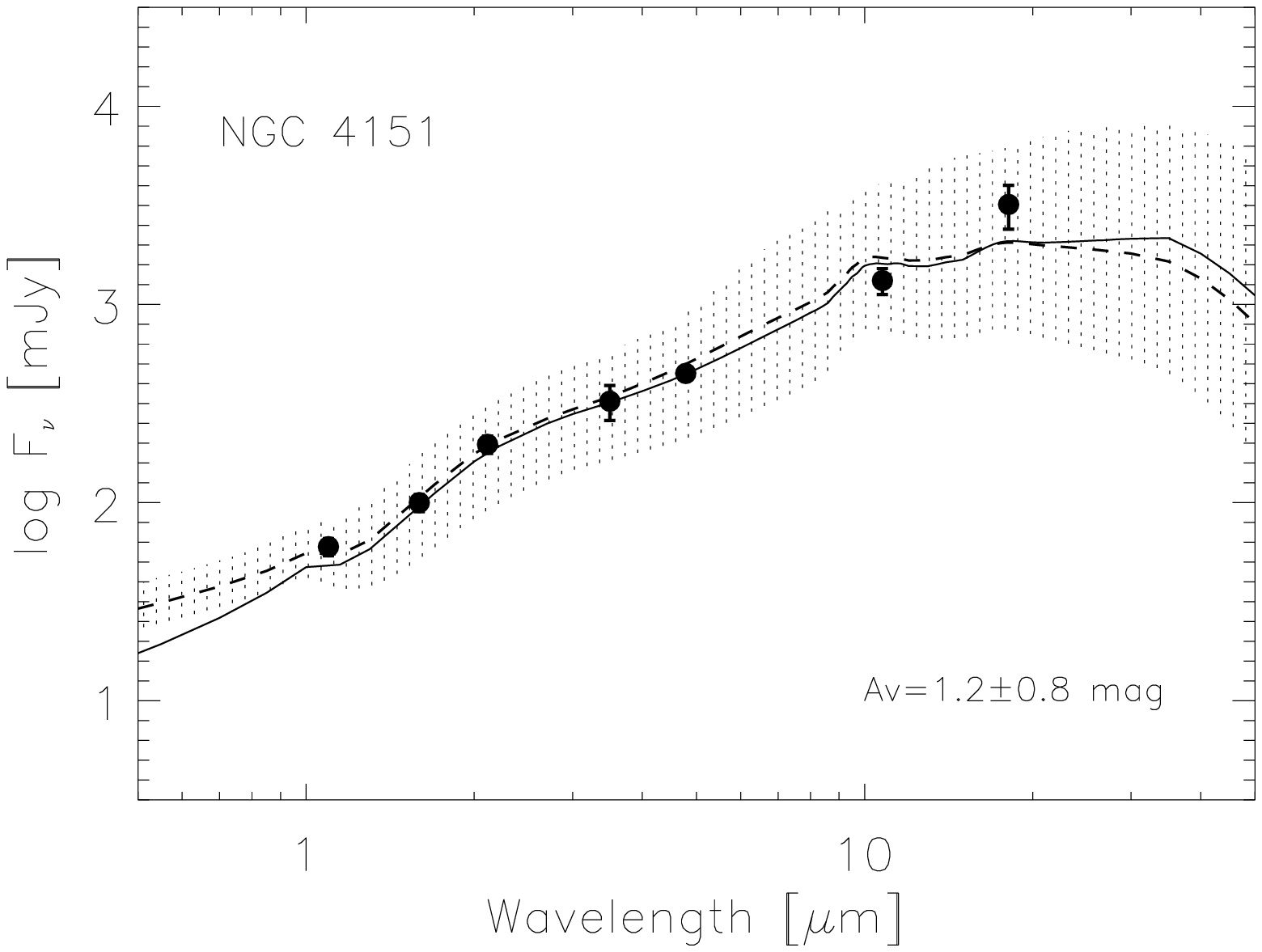}
\includegraphics[width=5.2cm]{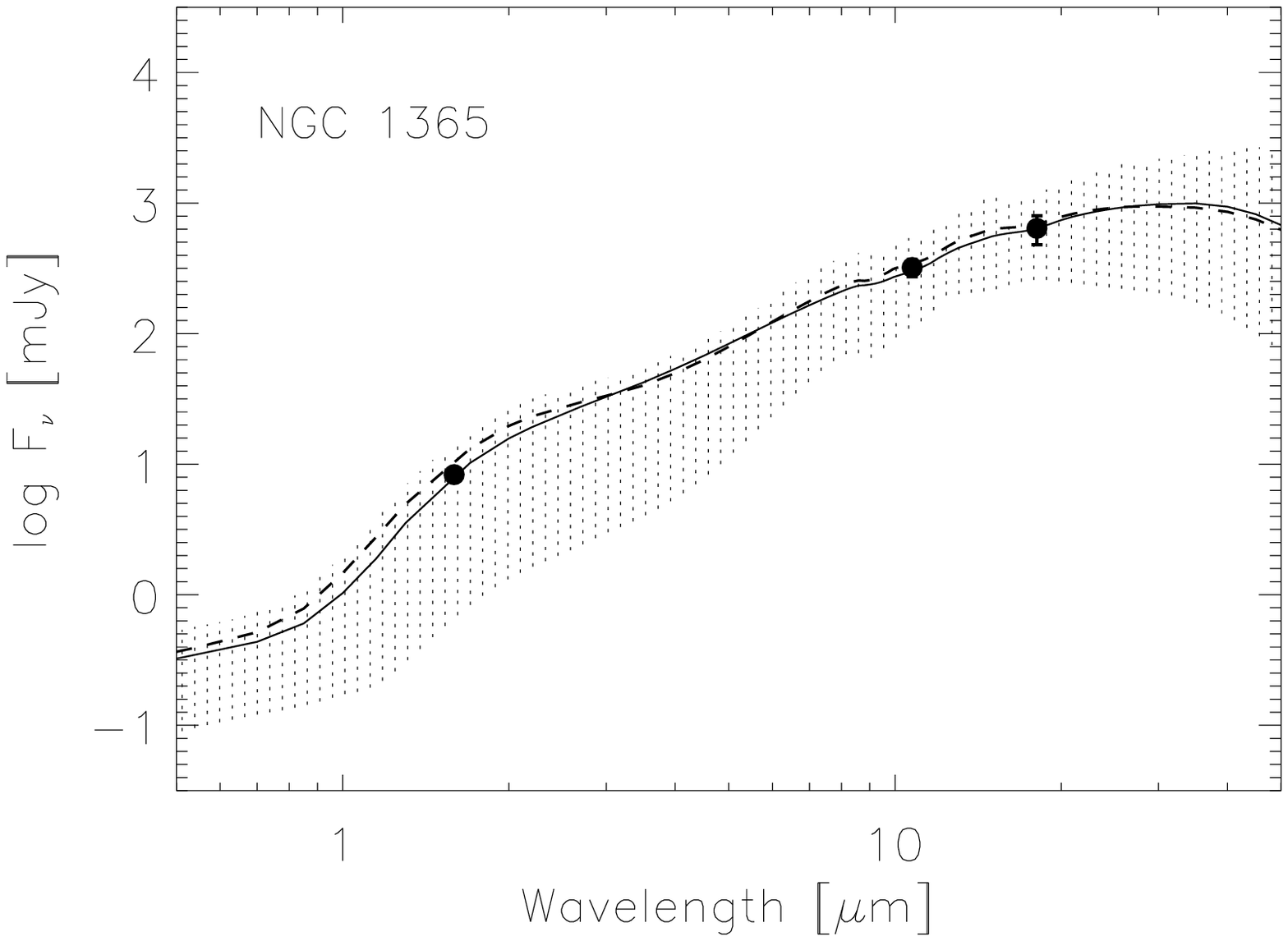}
\includegraphics[width=5.2cm]{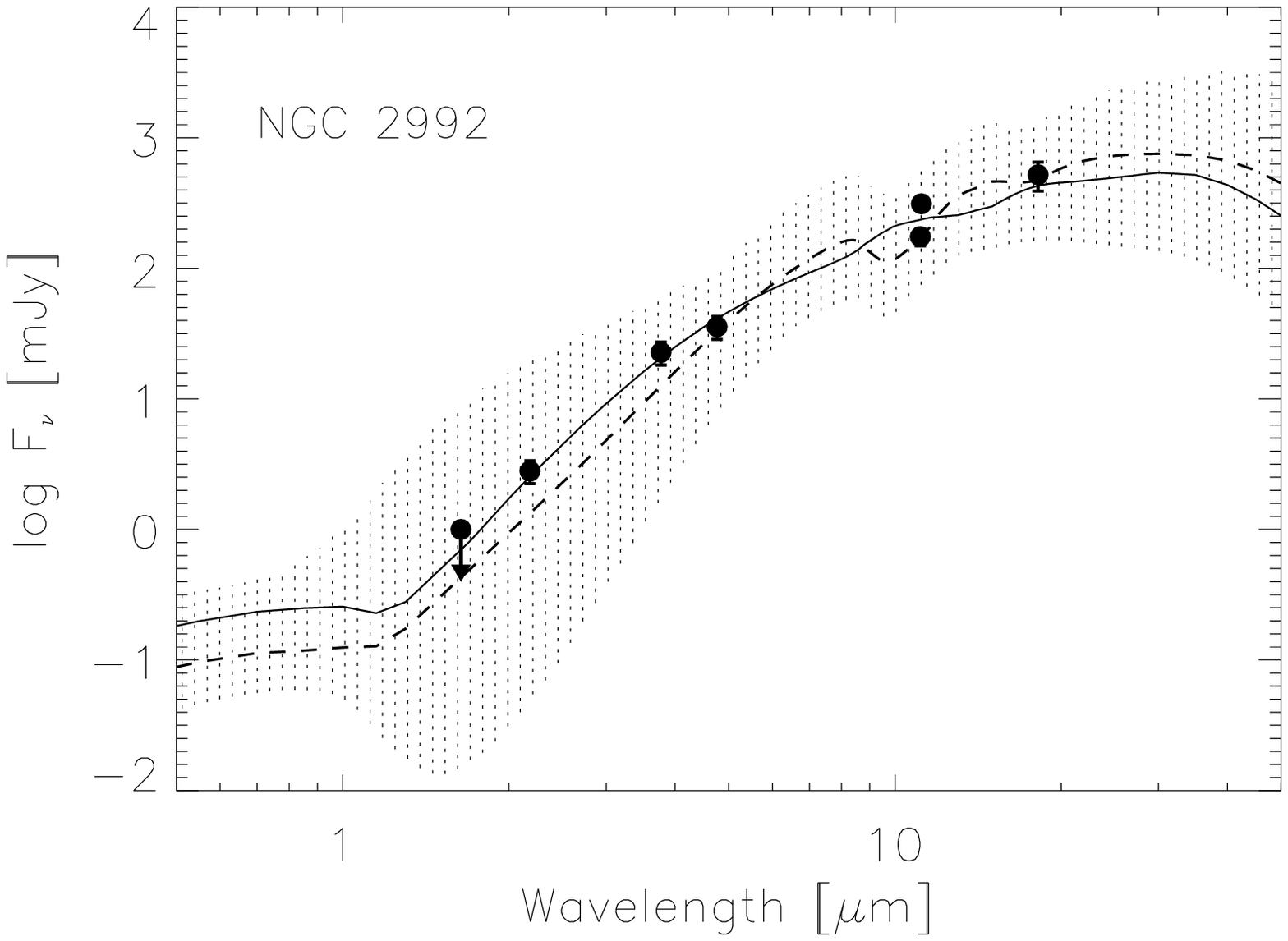}
\includegraphics[width=5.2cm]{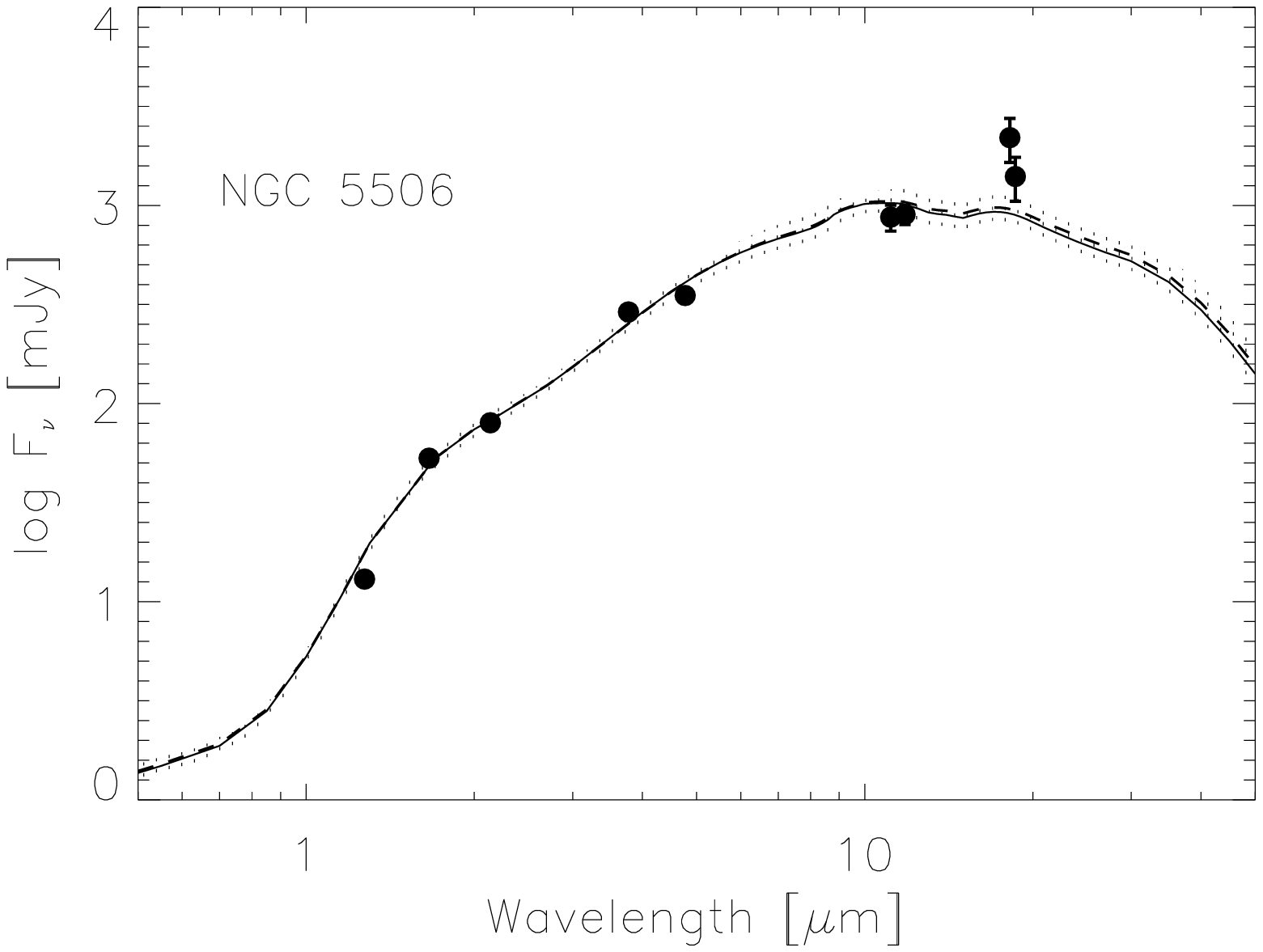}
\includegraphics[width=5.2cm]{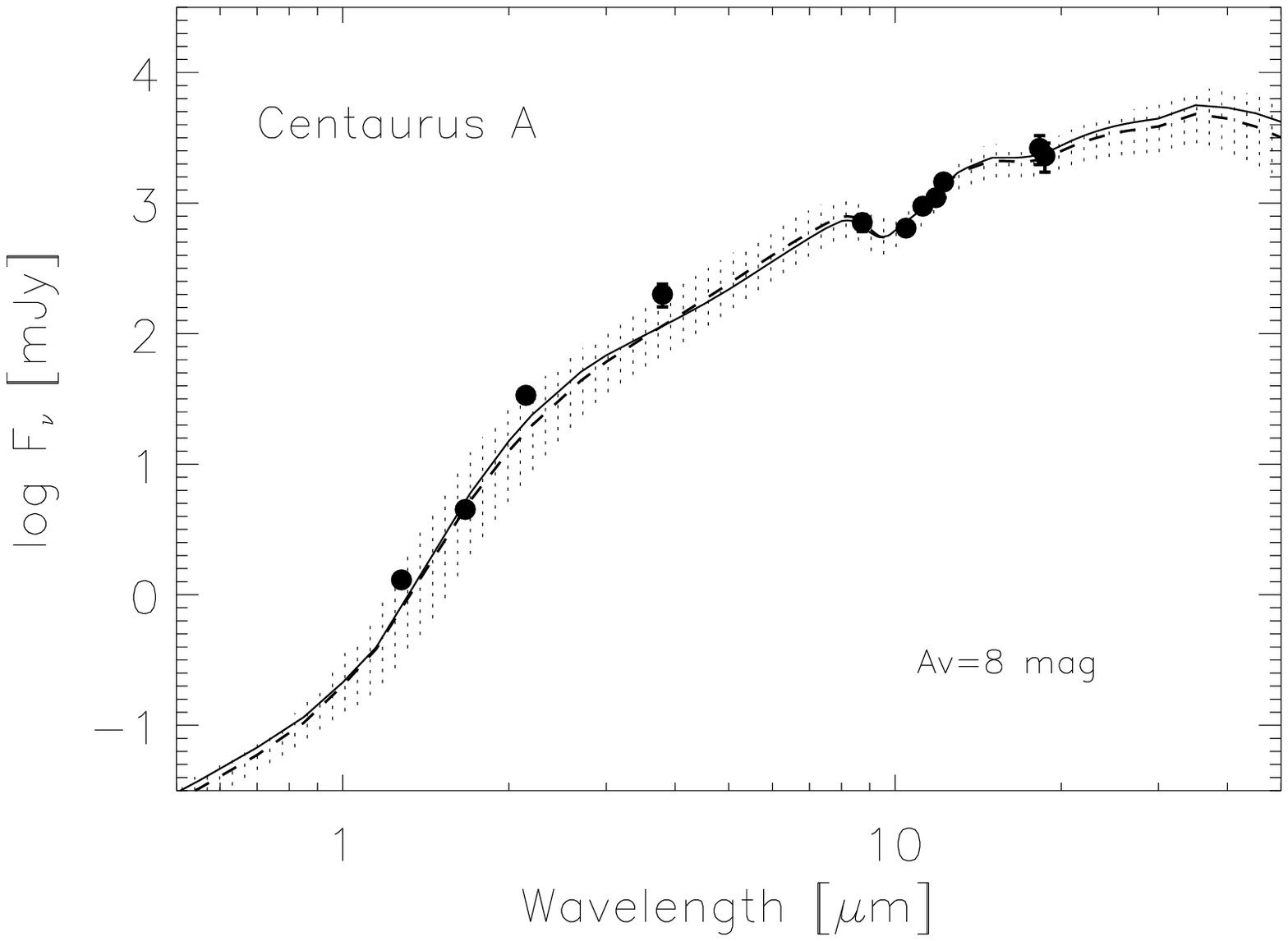}
\includegraphics[width=5.2cm]{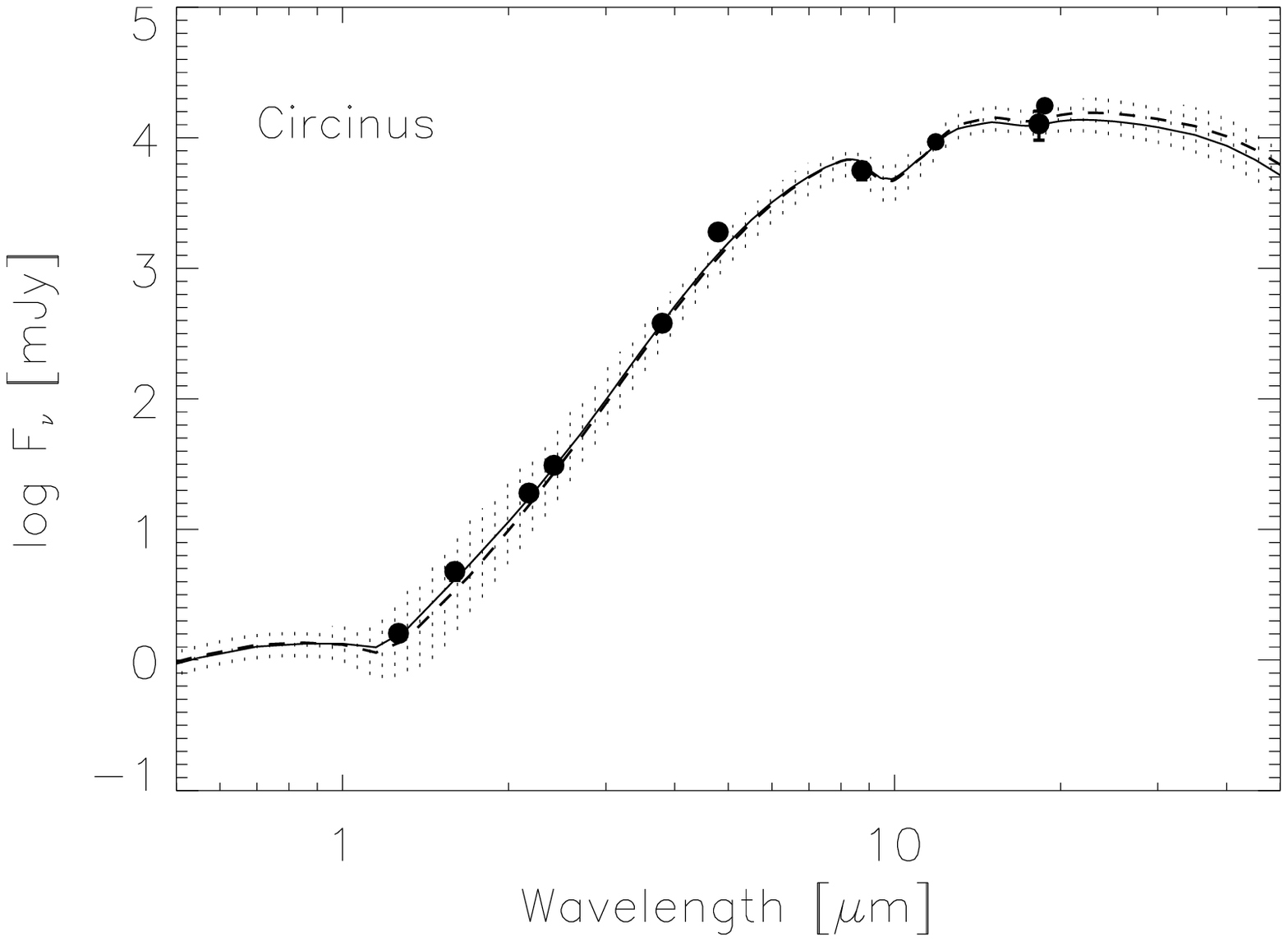}\par}
\caption{\footnotesize{High spatial resolution IR SEDs of the Seyfert galaxies (dots). 
Solid and dashed lines correspond 
to the MAP and median models, respectively. Shaded regions indicate the range of models compatible with the 
68\% confidence interval for each parameter around the median.}
\label{sy1_fits_a}}
\end{figure*}

\begin{figure*}[!ht]
\centering
{\par
\includegraphics[width=5.2cm]{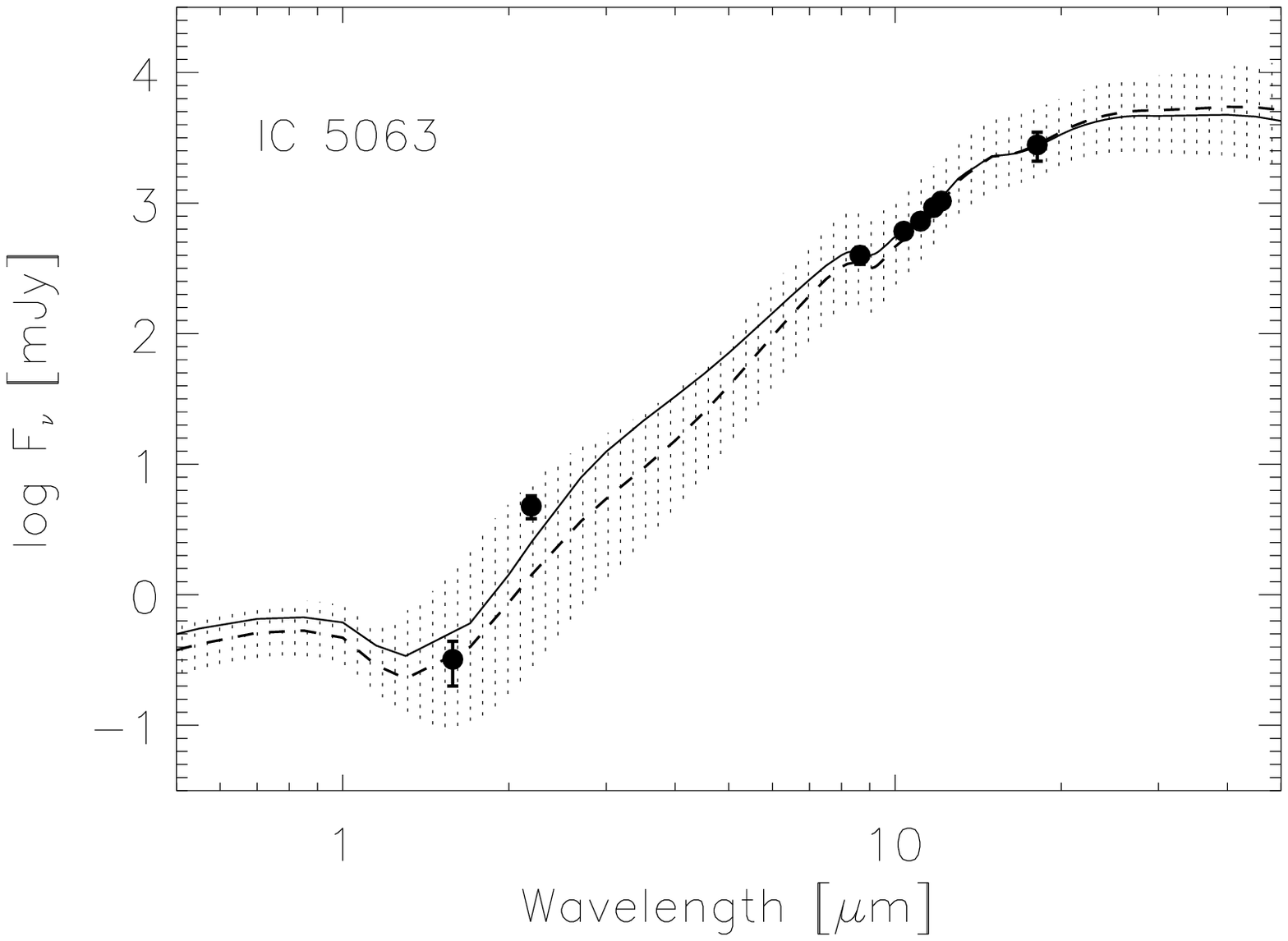}
\includegraphics[width=5.2cm]{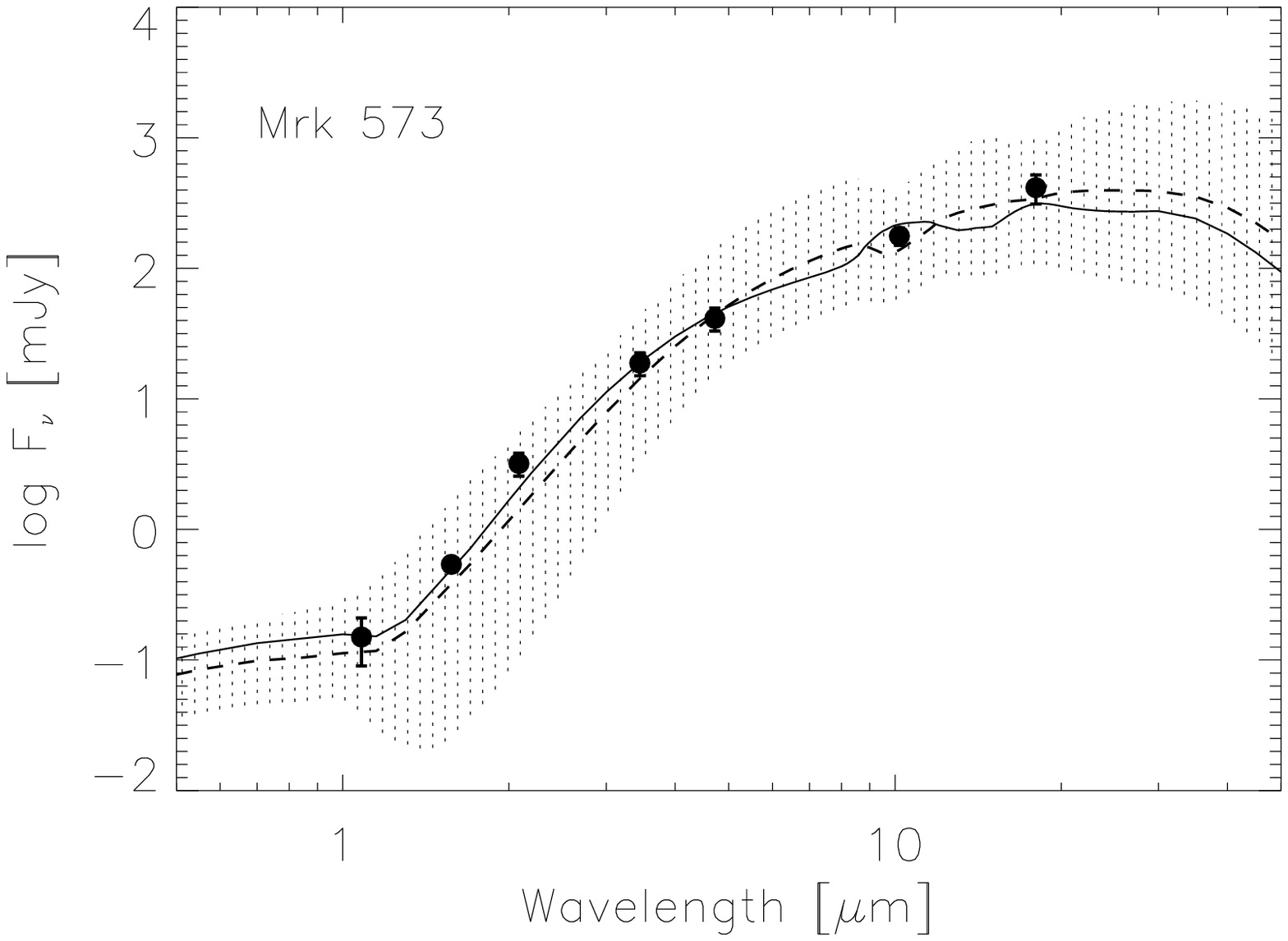}
\includegraphics[width=5.2cm]{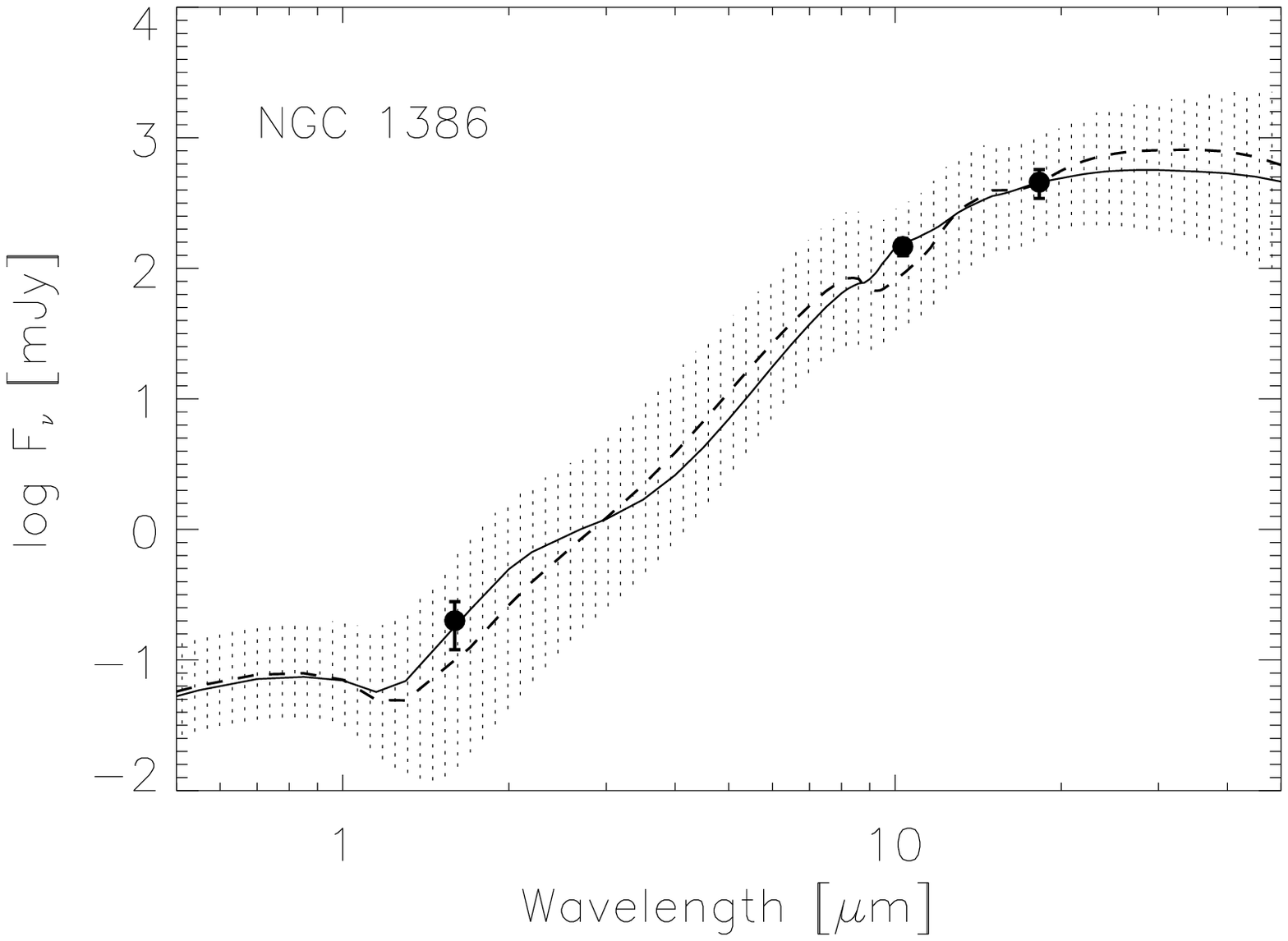}
\includegraphics[width=5.2cm]{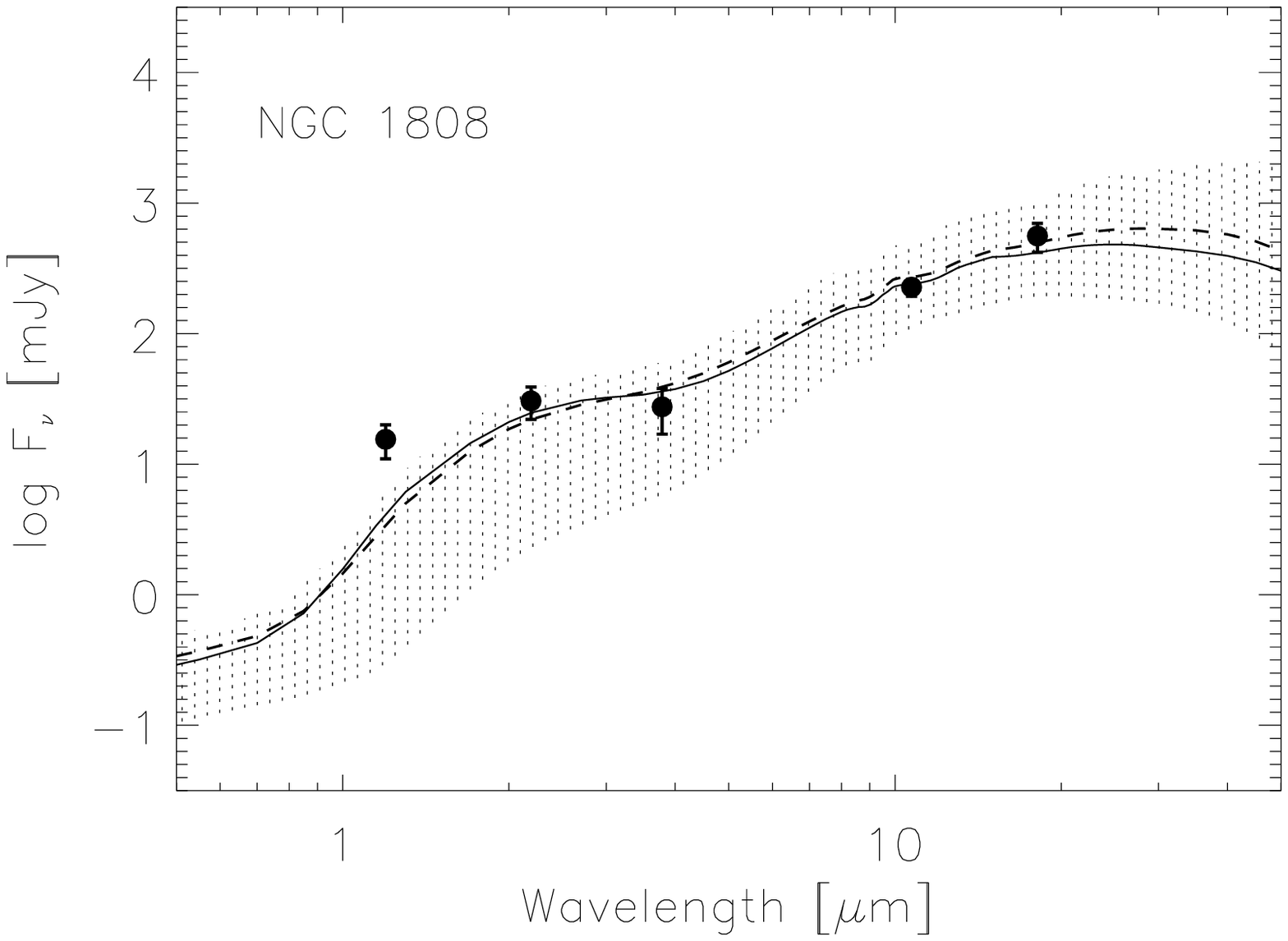}
\includegraphics[width=5.2cm]{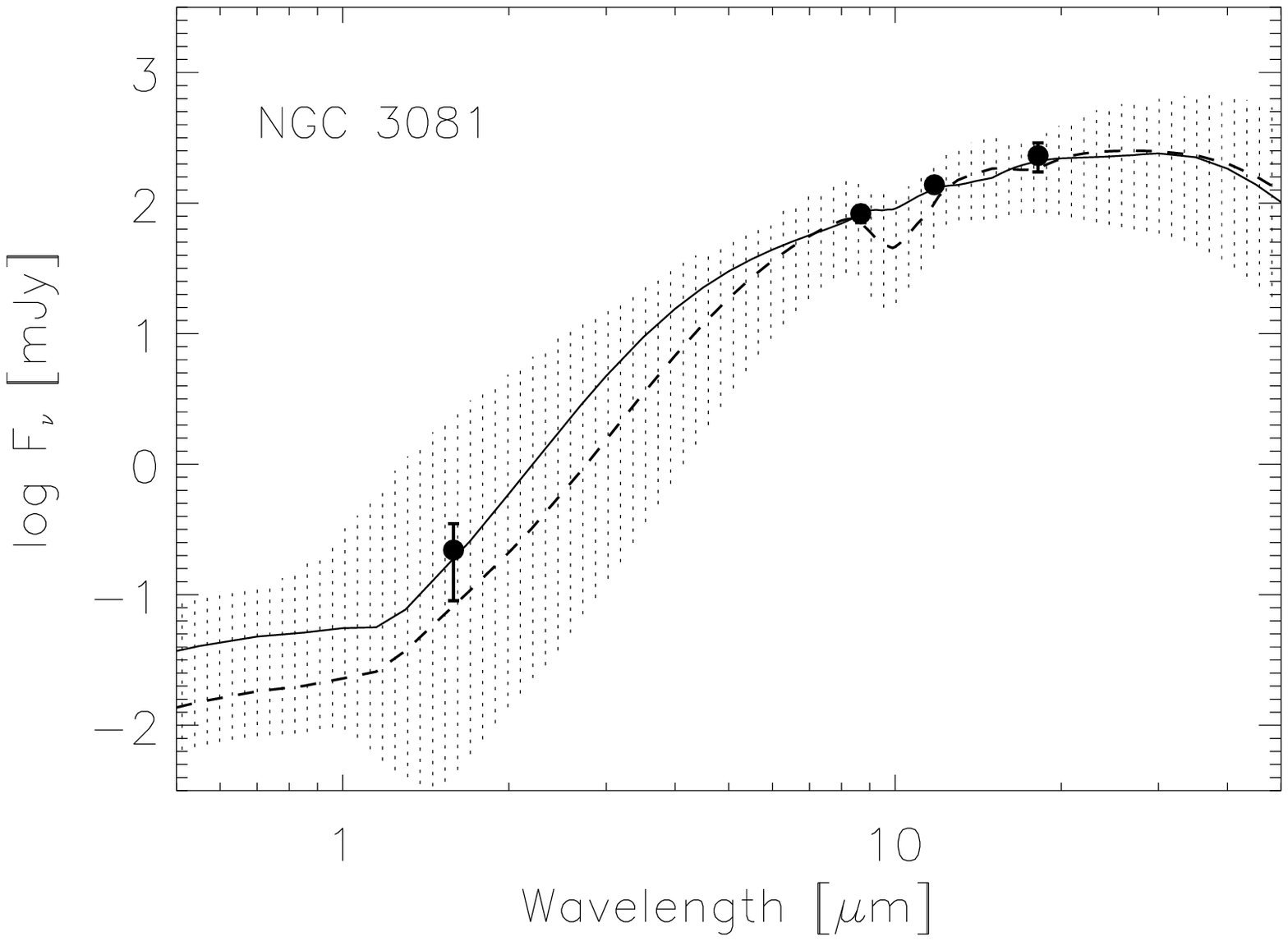}
\includegraphics[width=5.2cm]{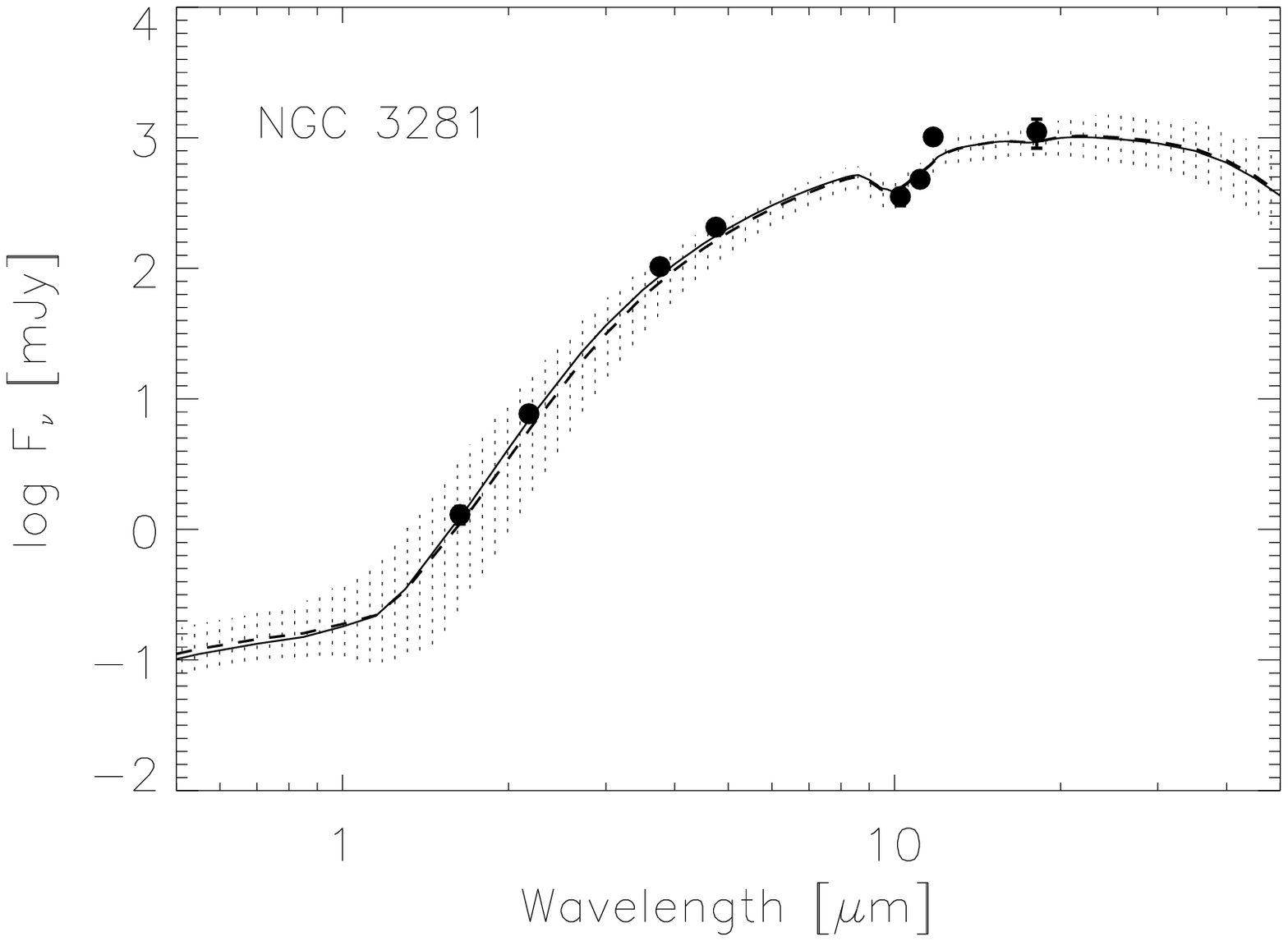}
\includegraphics[width=5.2cm]{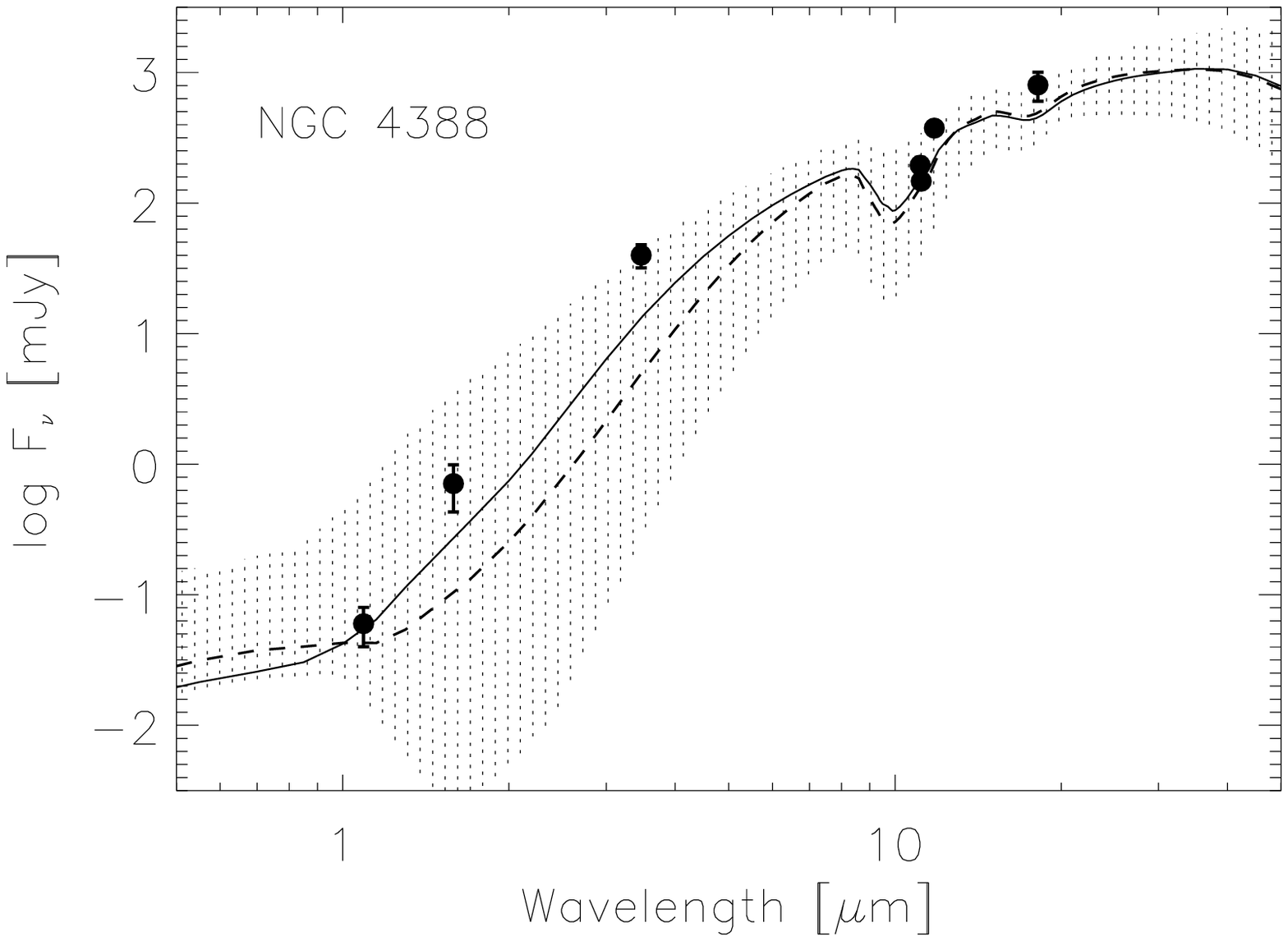}
\includegraphics[width=5.2cm]{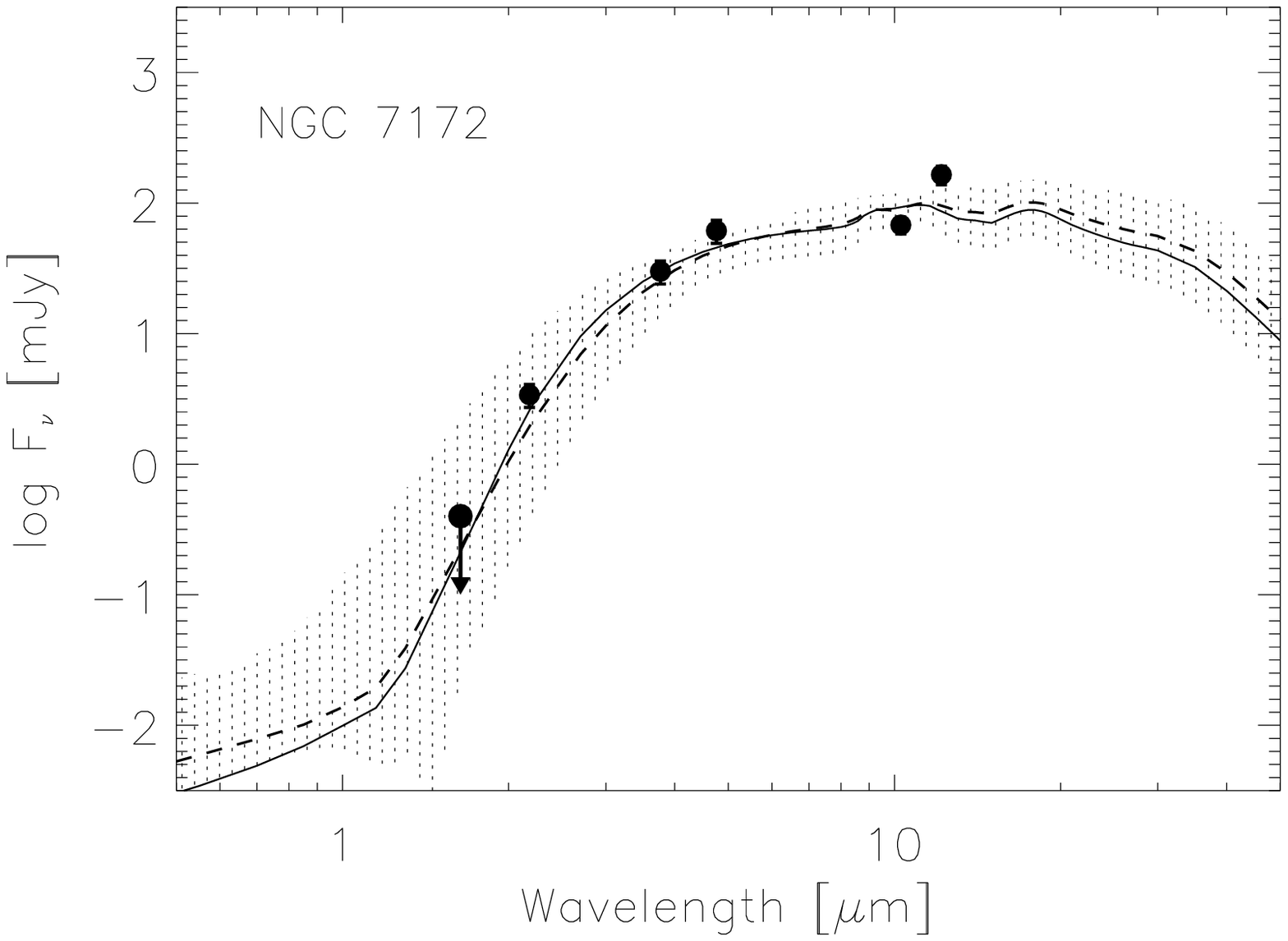}
\includegraphics[width=5.2cm]{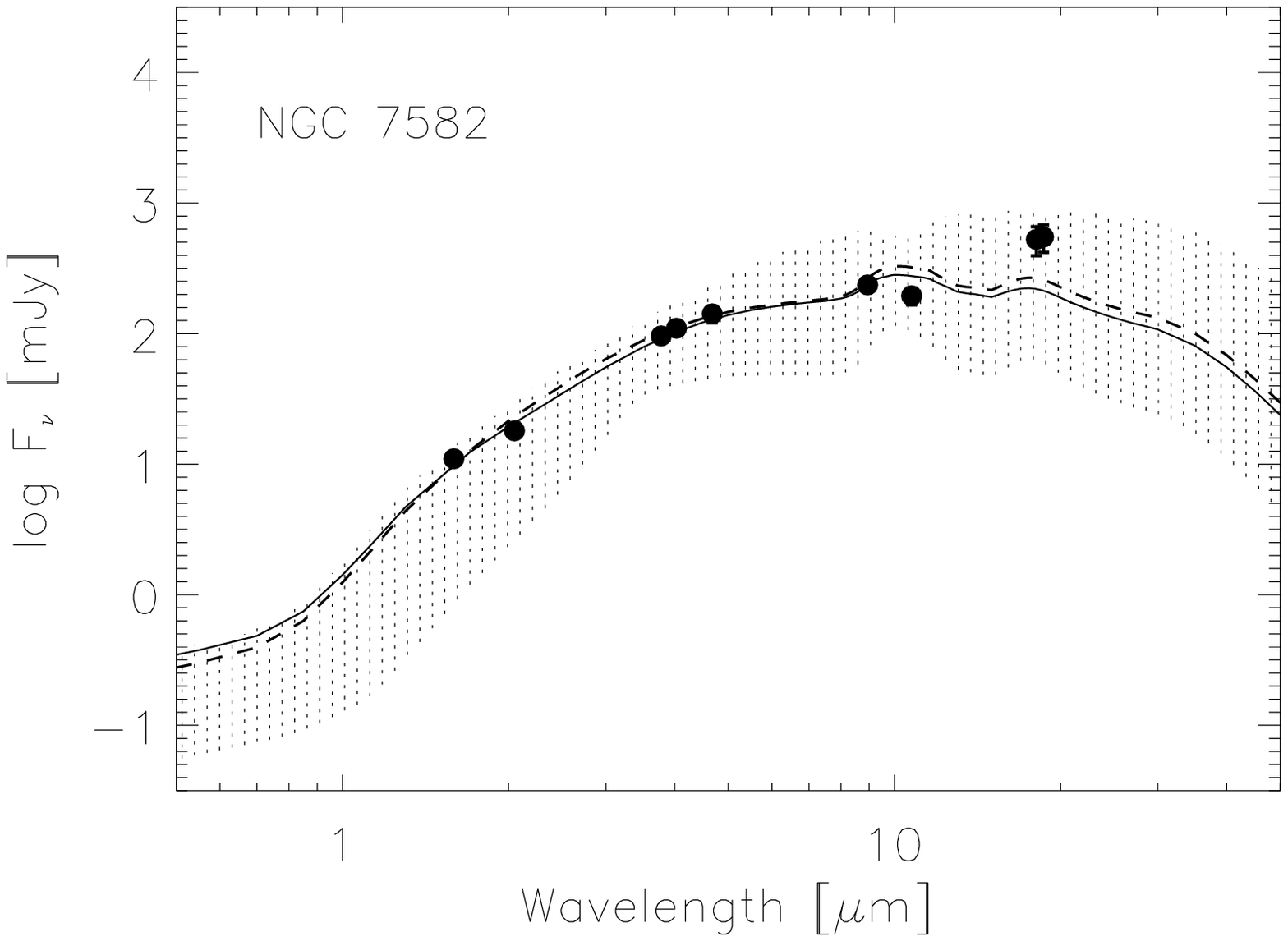}\par}
\caption{\footnotesize{Same as in Figure \ref{sy1_fits_a}, but for the remaining Sy2 galaxies.}
\label{sy1_fits_b}}
\end{figure*}

We fitted the individual IR SEDs with BayesClumpy,
modelling the torus emission and the direct AGN contribution (the latter as a broken power law).
We also consider the IR extinction curve of \cite{Chiar06} to take into account any possible foreground extinction 
from the host galaxy.
Although the solutions to the Bayesian inference problem are the
posterior distributions of each parameter, we can translate the results
into corresponding spectra (see Figures \ref{sy1_fits_a} and \ref{sy1_fits_b}).  
The solid lines correspond to the model described by 
the combination of parameters that maximizes their probability distributions (maximum-a-posteriori; MAP). 
Dashed lines represent the model computed with the median value of the probability distribution of 
each parameter. Shaded regions indicate the range of models compatible with the 
68\% confidence interval for each parameter around the median.

The clumpy models successfully reproduce the observed Seyfert SEDs studied here with compatible results
among them. This indicates that
the NIR and MIR unresolved fluxes employed here are dominated by a combination of 
reprocessed emission from dust in the torus and direct AGN emission.

\vspace{2cm}

\section{Comparison between Type 1 and Type 2 Seyfert tori.}
\label{discussion}

The aim of this work is to compare between the properties of Type 1 and 2 tori under 
the assumption that the SEDs studied here are torus/AGN dominated. Despite the 
relatively low number of objects considered (7 Sy1
and 9 Sy2; the unreliable fits of NGC 1808 and NGC 7582 are excluded), 
we find that some of the parameters are significantly different between Sy1 and Sy2.

\begin{figure*}[!ht]
\centering
\par{
\includegraphics[width=6cm]{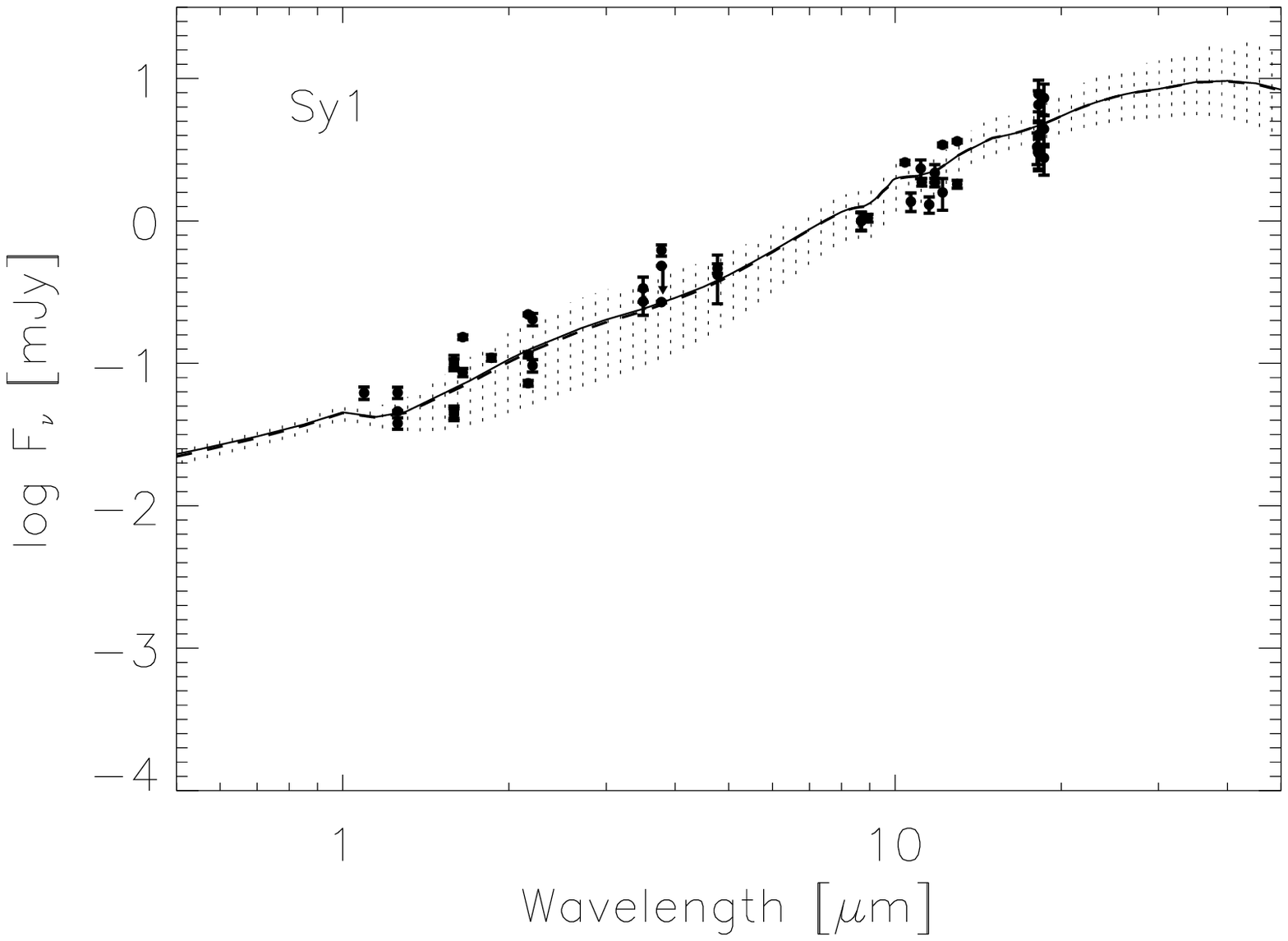}
\includegraphics[width=6cm]{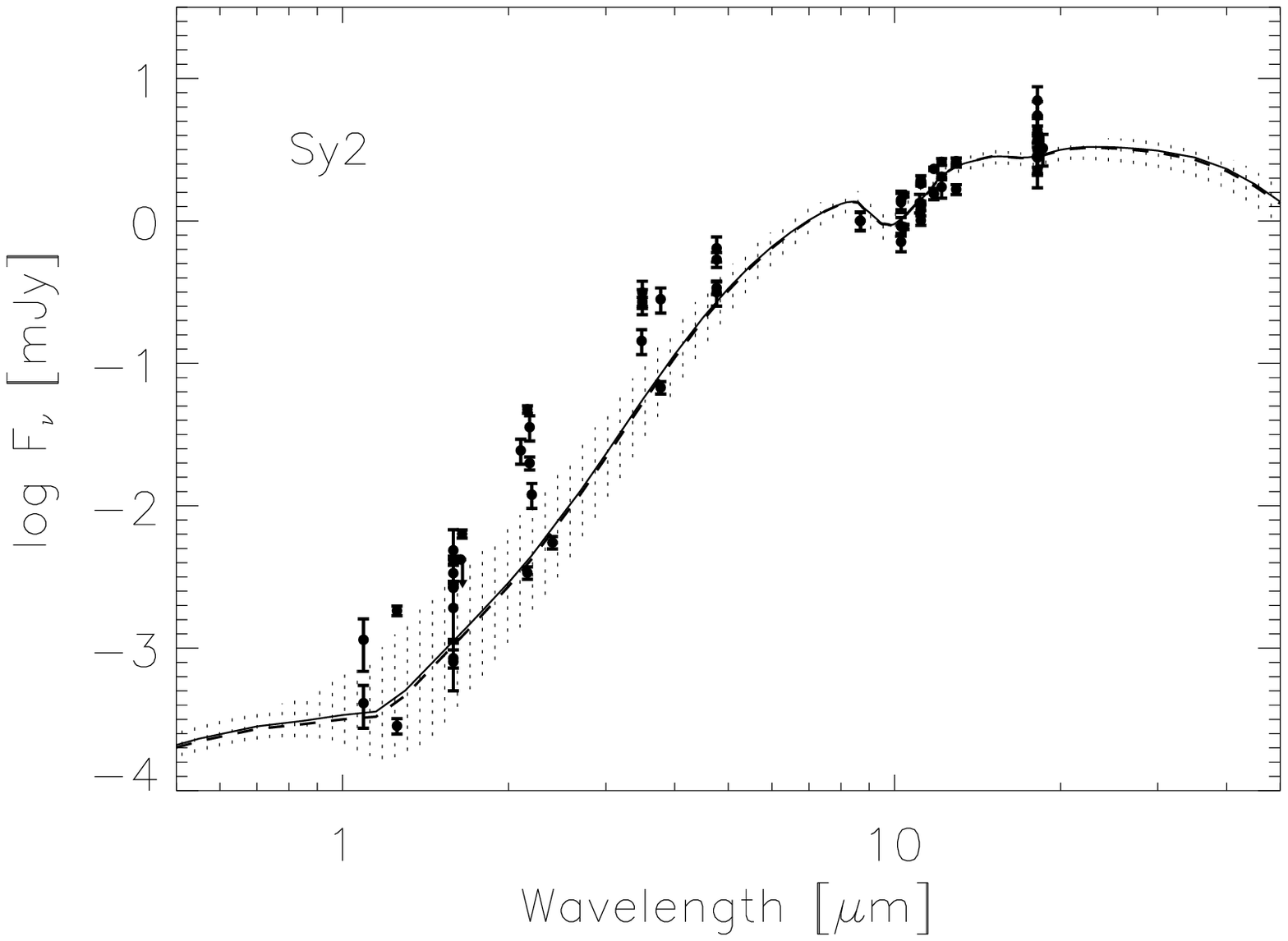}\par}
\caption{\footnotesize{Same as in Figure \ref{sy1_fits_a}, but for the Sy1 (left) and Sy2 SEDs (right) normalized at 8.74 $\mu$m.}
\label{comparison1and2}}
\end{figure*}

To take full advantage of the Bayesian approach, the best way to compare the
results for Sy1 and Sy2 galaxies is to derive joint posterior distributions for the full Type 1 and Type 2 datasets respectively.
Thus, we normalized all the Sy1 SEDs at 8.74 $\mu$m~and fitted them together using BayesClumpy, and we did the same for the Sy2. 
We considered the mean redshift for the Sy1 (z=0.0061$\pm$0.0045) and for the Sy2 (0.0078$\pm$0.0051) in the fits.    
In Figure \ref{comparison1and2} we show the Sy1 (left panel) and Sy2 fits (right panel). 
Note that the MAP and median models predict a flat SED with
the silicate feature in weak emission for the Sy1 galaxies, and steeper and with the silicate band in shallow absorption for Sy2. 

\begin{figure*}[!ht]
\centering
{\par
\includegraphics[width=5cm]{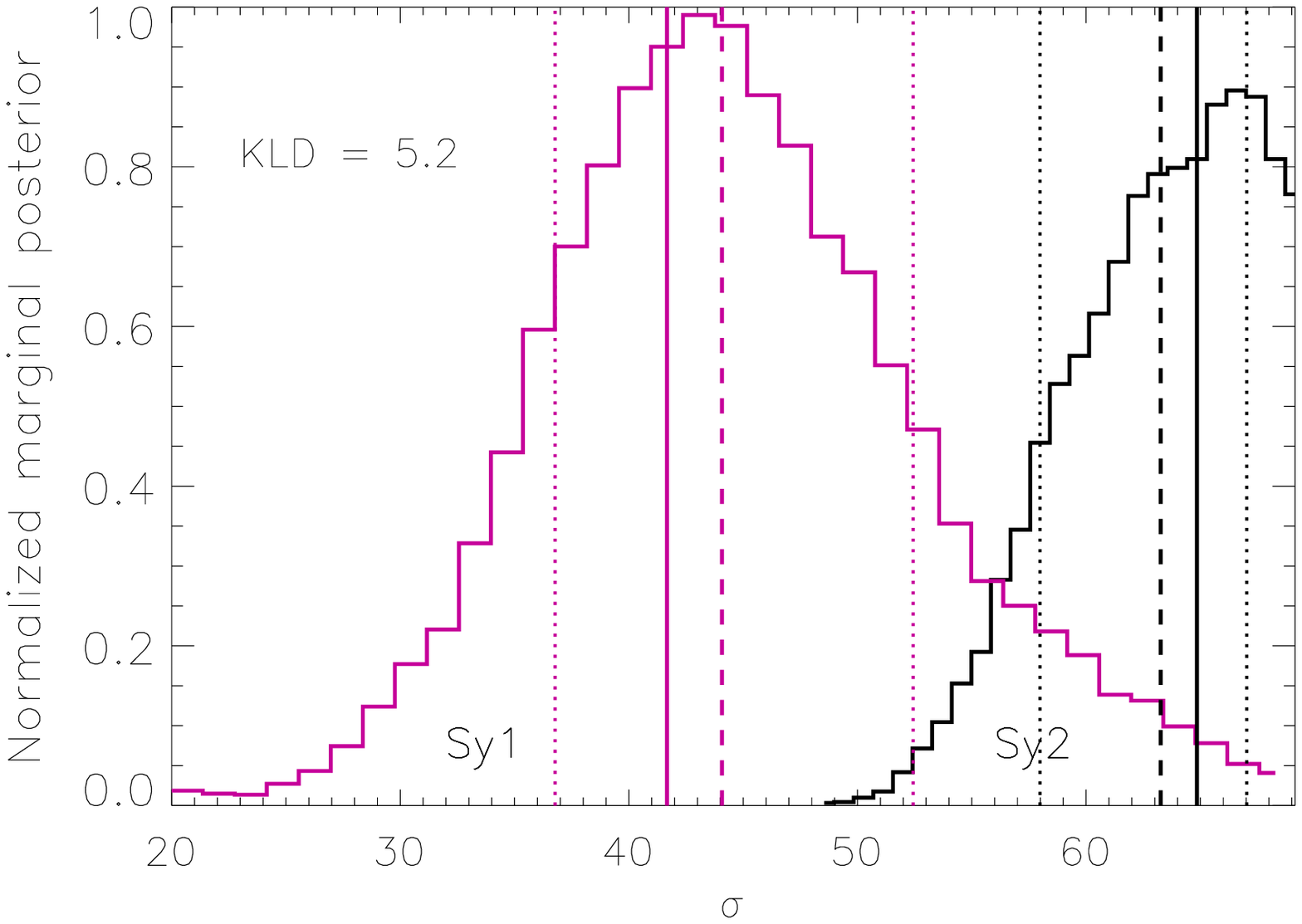}
\includegraphics[width=5cm]{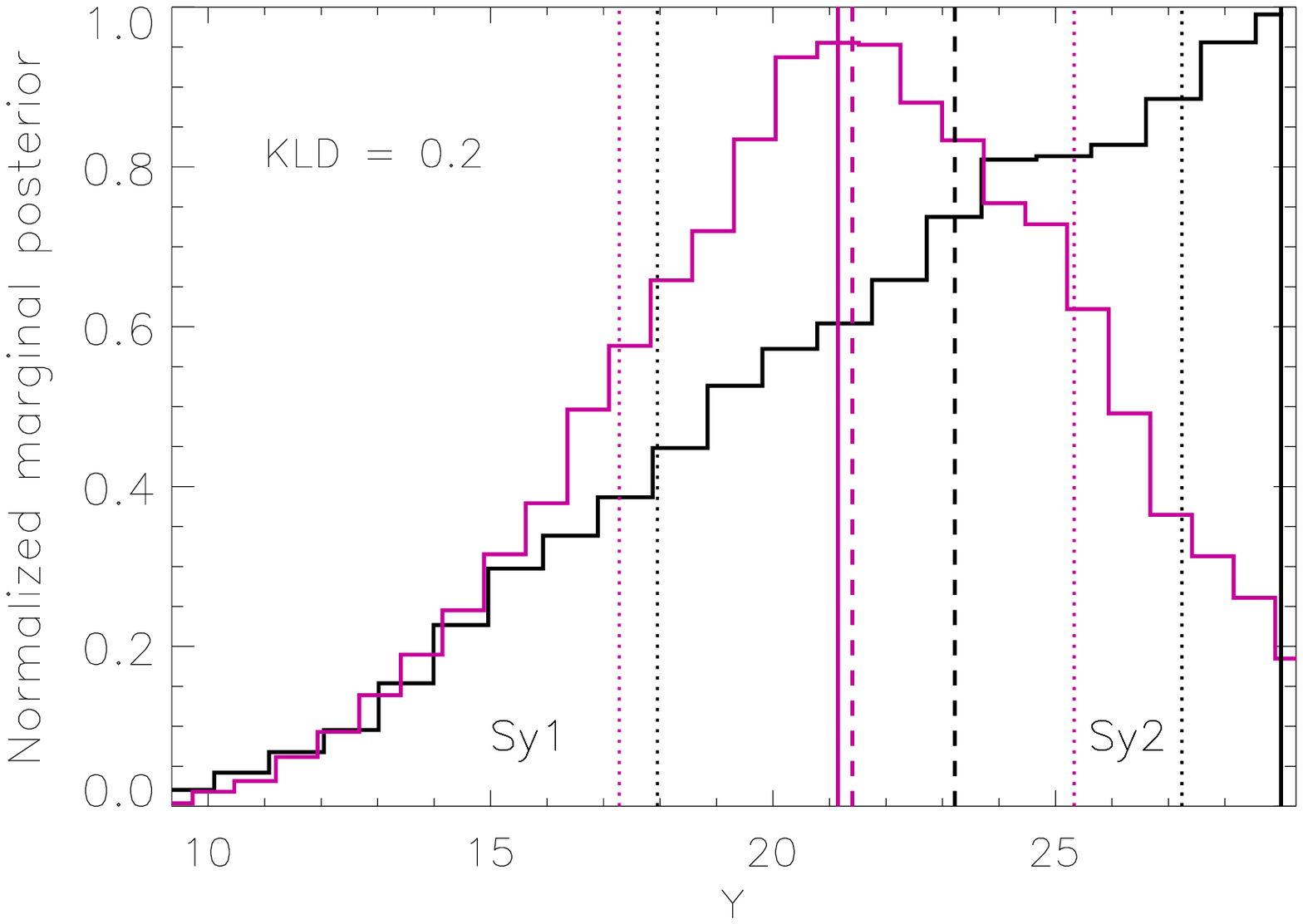}
\includegraphics[width=5cm]{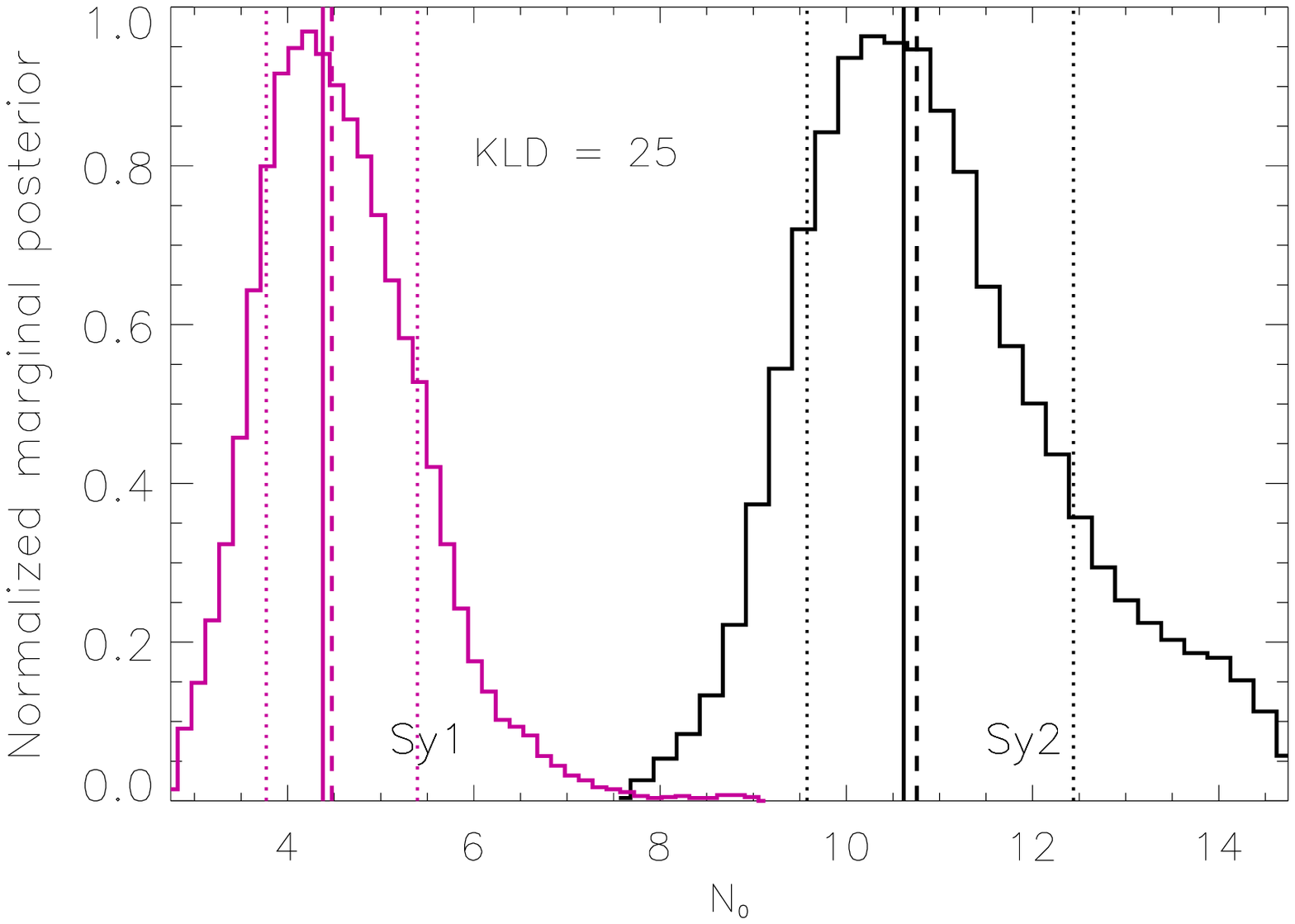}
\includegraphics[width=5cm]{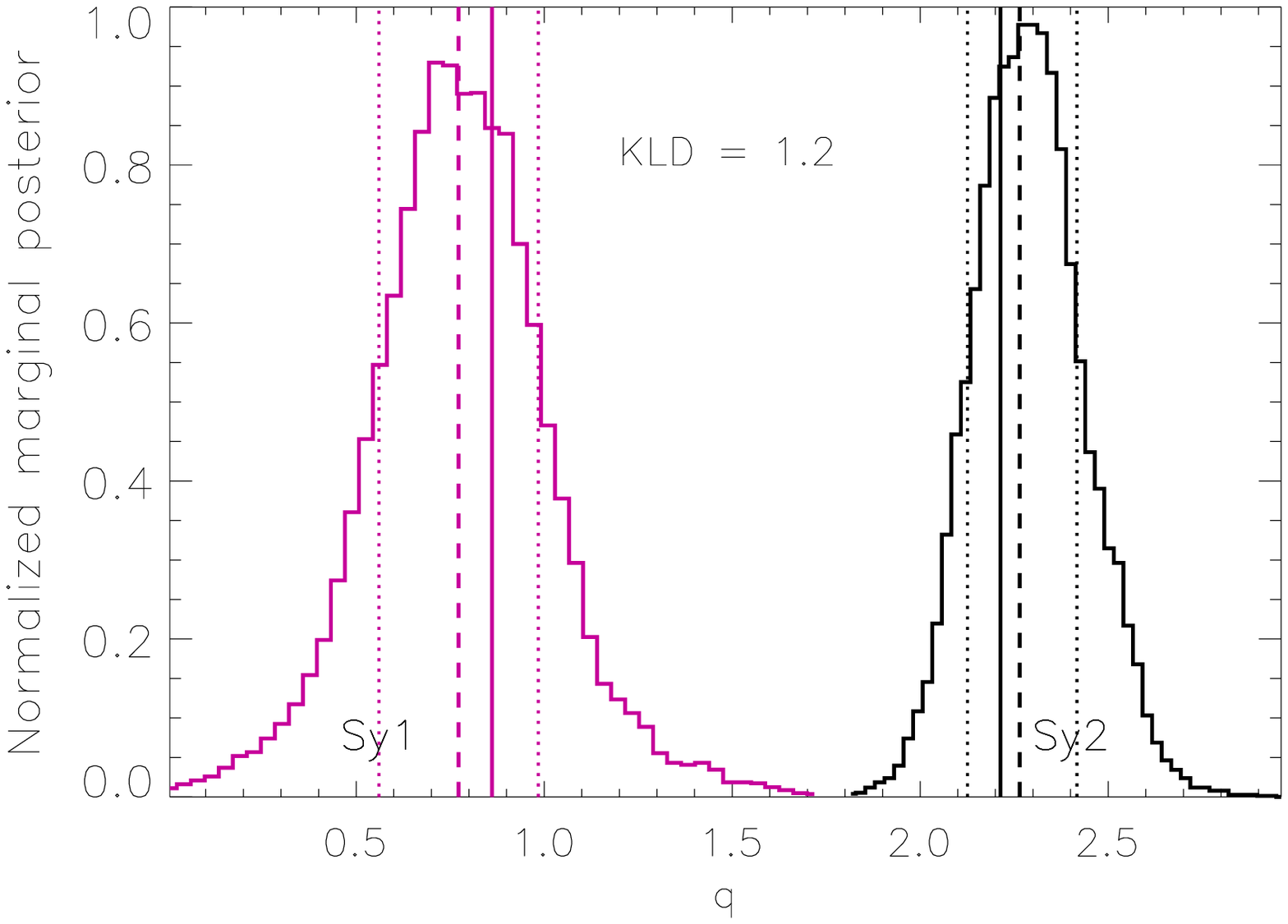}
\includegraphics[width=5cm]{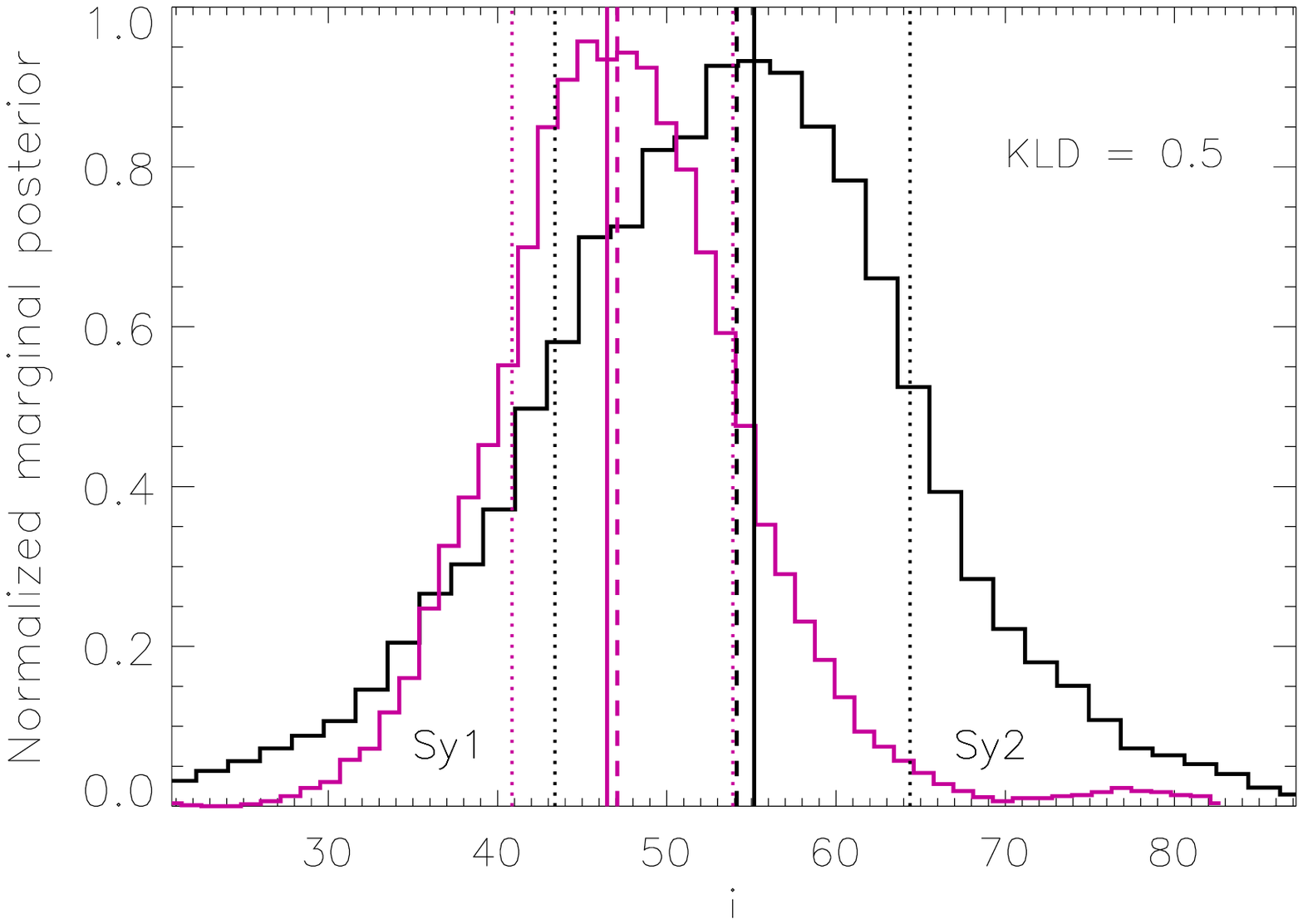}
\includegraphics[width=5cm]{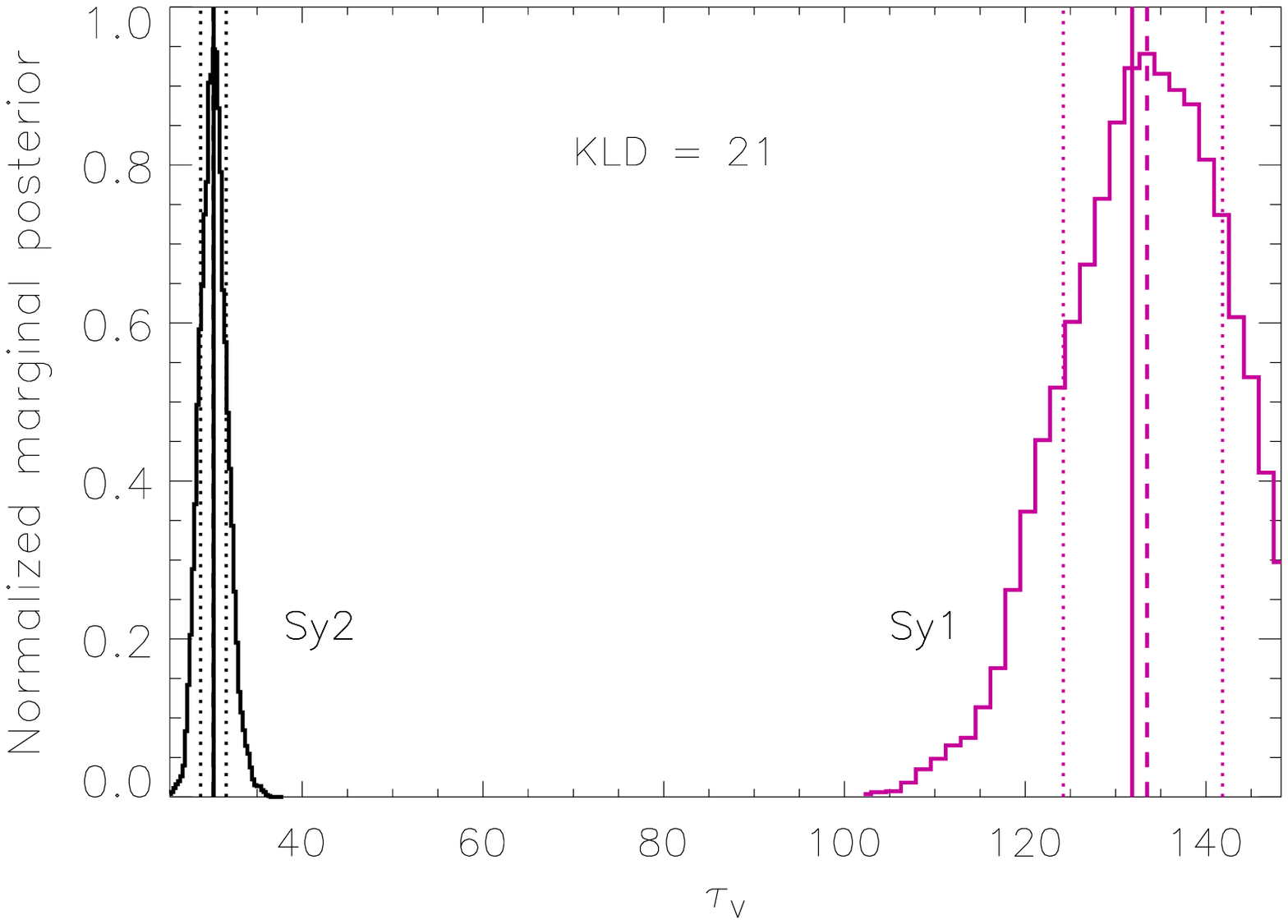}\par}
\caption{\footnotesize{Posterior distributions resulting from the fits of the joint Sy1 and Sy2 SEDs. 
KLD values derived from the comparison between Sy1 and Sy2 for each parameter are labelled.}
\label{divergence_plot}}
\end{figure*}

The comparison between the Sy1 and Sy2 posterior distributions is shown in Figure \ref{divergence_plot}. 
From a visual inspection it is clear that the joint posteriors of the parameters N$_0$, $q$, $\tau_V$, and $\sigma$
are completely different between Sy1 and Sy2. There is no overlap between the 
1-sigma intervals. 
In Table 3 we report the median and mode values of the histograms in Figure \ref{divergence_plot}.

In order to quantify how different the probability distributions are, we calculated the 
Kullback-Leibler divergence (KLD; \cite{Kullback51}) between the Sy1 and Sy2 posteriors. This divergence takes into
account the full shape of the posterior and it is
a positive value, and equal to zero when two distributions are identical. Therefore, the larger the value of KLD, 
the more different the posteriors. We find KLD$>$1 for $\sigma$, $N_0$, $q$, and $\tau_V$
(see Figure \ref{divergence_plot}). These are indeed the four parameters whose 1-sigma regions do not overlap.
Thus, we consider their differences significant between Sy1 and Sy2. 
For both $Y$ and $i$ we find KLD$<$1 and similar median values between Sy1 and Sy2.

\begin{table}
\caption{Statistics of the comparison between Sy1 and Sy2 parameters}
\begin{footnotesize}
\begin{tabular}{lllllllllllll}
\br
Type & \multicolumn{2}{c}{$\sigma$ (deg)} & \multicolumn{2}{c}{$Y$} &  \multicolumn{2}{c}{$N_0$} & \multicolumn{2}{c}{$q$} & \multicolumn{2}{c}{$i$ (deg)} & \multicolumn{2}{c}{$\tau_{V}$} \\
 & Sy1 & Sy2  & Sy1 & Sy2   & Sy1 & Sy2 & Sy1 & Sy2  & Sy1 & Sy2 &  Sy1 & Sy2 \\
\mr
Medians & 44$\pm^{8}_{7}$ & 63$\pm^{4}_{5}$ & 21$\pm$4 & 23$\pm^{4}_{5}$ & 4$\pm$1 & 11$\pm^{2}_{1}$ & 0.8$\pm$0.2 & 2.3$\pm$0.1 & 47$\pm^{7}_{6}$ & 54$\pm^{10}_{11}$  & 133$\pm^{8}_{9}$ & 30$\pm$1	\\
Modes   & 42              & 65              & 21       & 29	         & 4       & 11 	     & 0.9	   & 2.2         & 46		   & 55 		& 132		   & 30 	\\  
\mr          
\end{tabular}
\end{footnotesize}
\end{table}

Sy1 tori are narrower and have fewer clouds ($\sigma=44^o\pm^{8^o}_{7^o}$; N$_0$=4$\pm$1) than those of Sy2 
($\sigma=63^o\pm^{4^o}_{5^o}$; N$_0$=11$\pm^{2}_{1}$). The radial density distribution of the clouds is also 
different between the two Seyfert types according to this analysis: in Sy2, the majority of the clumps are distributed very close to the nucleus 
(i.e. steep radial density distribution; $q$=2.3$\pm$0.1) whereas for Sy1 the clouds distribution is flatter ($q$=0.8$\pm$0.2).
On the other hand, the optical depth of the clouds in Sy1 tori is larger ($\tau_V$=133$\pm^{8}_{9}$) than in 
Sy2 ($\tau_V$=30$\pm$1). 

Interestingly, we find high as well as low values of the inclination angle of the torus for Sy1 and Sy2 (see Table 3). 
This variety in the $i$ values translates into the similar median values of the joint Sy1 and Sy2 posteriors
(47$^o\pm^{7^o}_{6^o}$ for Sy1 and 54$^o\pm^{10^o}_{11^o}$ for Sy2), which are intermediate within the considered
prior ($i$=[0$^o$,90$^o$]). {\it This is telling us that, in the clumpy torus scenario, the classification of a 
Seyfert galaxy as a Type 1 or 2 may depend on the intrinsic properties of the torus rather than in mere inclination.}

\begin{figure*}[!ht]
\centering
\includegraphics[width=10cm]{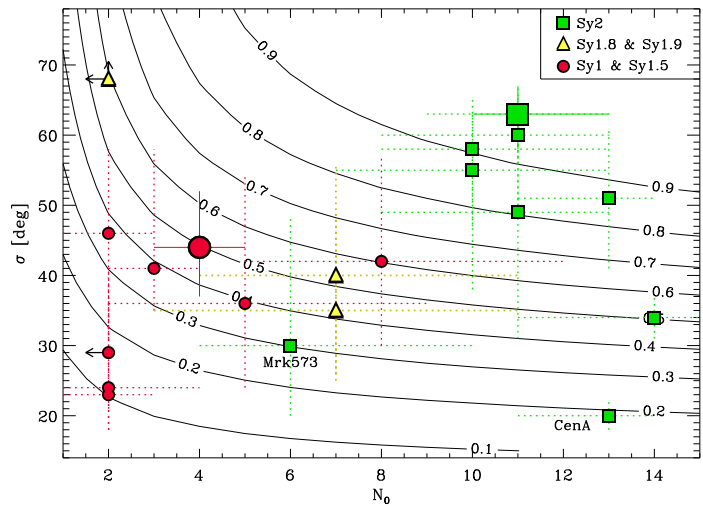}
\caption{\footnotesize{$\sigma$ versus N$_0$ for the individual galaxies. Either median values or upper/lower limits
are taken from the fits presented here. 
Error bars indicate 68\% confidence level around the median.
The big dot and square correspond to the average $\sigma$ and N$_0$
values for Sy1 and Sy2 from Table 3.
\label{covering}}}
\end{figure*}

In Figure \ref{covering} we represent the median values of $\sigma$ and $N_0$ for the different 
Seyfert types over the covering factor contours. The covering factor is defined as C$_T=1-\int e^{-N_{LOS}(i)}d cos(i)$.
Type 1 nuclei tend to be located within lower C$_T$ contours (C$_T\leq$0.6)
than those of Type 2s, for which C$_T\geq$0.5, with the exceptions of Centaurus A and Mrk 573.
We have represented with larger symbols the median values from the joint $\sigma$ and $N_0$ posteriors 
reported in Table 3. 

Since the covering factor is a non-linear function of the torus parameters, we took full advantage of our
Bayesian approach and generated joint posterior distributions for C$_T$ from those in Figure  \ref{divergence_plot}. 
The median values of the histograms are C$_T$(Sy2)=0.95$\pm$0.02 
and C$_T$(Sy1)=0.5$\pm$0.1. The divergence between the Sy1 and Sy2 C$_T$ posteriors is KLD=28, indicating that the difference 
is significant (1-sigma regions do not overlap). Thus, Sy1 tori in our sample have lower C$_T$s than those of Sy2, implying that
they are intrinsically different. 
 
The clumpy model fits yield the intrinsic bolometric luminosity of 
AGN (L$_{bol}^{AGN}$) by means of the vertical shift applied to match the observational data points.
Combining this value with the torus luminosity (L$_{bol}^{tor}$), obtained by integrating the corresponding model torus emission 
(without the AGN contribution), we derive the reprocessing efficiency (RE) of the torus (L$_{bol}^{tor}$/$L_{bol}^{AGN}$).
The previous values are calculated on the Bayesian framework, by combining the posterior distributions of the 
model parameters. 
Sy2 tori in our sample are more efficient reprocessors than Sy1, absorbing and re-emitting the
majority of the intrinsic AGN luminosity in the IR: RE(Sy2)=[0.4, 1.0], with a median value of 0.8 and 
RE(Sy1)=[0.2,0.7], with median of 0.5.

We considered a possible dependency of the RE (or alternatively the covering factor; C$_T$) on L$_{bol}^{AGN}$, since 
the amount of incoming radiation from the AGN could possibly have some influence on the reprocessed energy or 
even in the torus properties (e.g., receeding torus scenario; \cite{Lawrence91}). 
However, we find no relationship between the two quantities in the luminosity range considered. 
This means that the reprocessing efficiency 
depends primarily on the total number of clouds available to absorb the incident radiation, i.e. on the torus covering factor. 
However, the possible dependence of the torus properties on the AGN luminosity considering a broader luminosity range is further investigated
in \cite{Alonso11}.

In general, the IR SED fitting does not constrain the size of the torus ($Y$) as well as other 
model parameters (see Section 5.3 in \cite{Ramos09a}). The NIR and MIR observations are sensitive
to the warm dust (located within $\sim$10 pc of the nucleus), which depends on the combination 
of model parameters N$_0$, $q$, and $Y$ \cite{Thompson09}. Far-IR observations are more sensitive
to the torus extent independently, as shown in \cite{Ramos11b}.
The outer size of the torus scales with the AGN bolometric luminosity:
$R_{o} = Y R_{d}$, so assuming a dust sublimation temperature of 1500
K, $R_o= 0.4~Y~(L_{bol}^{AGN}/10^{45})^{0.5}$ pc. We derived 
$R_o$ posterior distributions from those of L$_{bol}^{AGN}$ and $Y$ and 
find that all tori in our sample have outer radii smaller than 6 pc, 
in agreement with MIR direct imaging of nearby Seyferts \cite{Packham05,Radomski08} and 
also interferometric observations \cite{Jaffe04,Tristram07,Meisenheimer07,Raban09}.

{\it Summarizing, we find tantalizing evidence, albeit for a small sample of Seyfert galaxies and under the clumpy torus hypotesis, that the classification as a Type 1 and Type 2 depends on the intrinsic torus properties, in contradiction with the simplest unification model.}


\section*{References}
\begin{thebibliography}{9}


\bibitem{Alonso11} Alonso-Herrero, A., et al. 2011, ApJ, 736, 82
	
\bibitem{Alonso03} Alonso-Herrero, A., Quillen, A. C., Rieke, G. H., Ivanov, V. D., 
\& Efstathiou, A. 2003, AJ, 126, 81

\bibitem{Alonso01} Alonso-Herrero, A., Quillen, A. C., Simpson, C, Efstathiou, A., \& Ward, M. J.
2001, AJ, 121, 1369

\bibitem{Antonucci93} Antonucci, R. R. J. 1993, ARA\&A, 31, 473

\bibitem{Asensio09} Asensio Ramos, A. \& Ramos Almeida, C. 2009, ApJ, 696, 2075

\bibitem{Barvainis87} Barvainis, R. 1987, ApJ, 320, 537

\bibitem{Chiar06} Chiar, J. E. \& Tielens, A. G. G. M. 2006, ApJ, 637, 774

\bibitem{Efstathiou95} Efstathiou, A., \& Rowan-Robinson, M. 1995, MNRAS, 273, 649

\bibitem{Fadda98} Fadda, D., Giuricin, G., Granato, G. L., \& Vecchies, D. 1998, ApJ, 496, 117

\bibitem{Granato94} Granato, G. L., \& Danese, L. 1994, MNRAS, 268, 235 

\bibitem{Granato97} Granato, G. L.,  Danese, L., \& Franceschini, A. 1997, ApJ, 486, 147
	
\bibitem{Honig10} H\"onig, S. F., et al. 2010, A\&A, 515, 23

\bibitem{Honig06} H\"onig, S. F., Beckert, T., Ohnaka, K., \& Weigelt, G. 2006, A\&A, 452, 459

\bibitem{Horst09} Horst, H., Duschl, W. J., Gandhi, P., \& Smette, A. 2009, A\&A, 495, 137

\bibitem{Jaffe04} Jaffe, W., et al. 2004, Nature, 429, 47

\bibitem{Krolik88} Krolik, J. H. \& Begelman, M. C. 1988, ApJ, 329, 702

\bibitem{Kullback51} Kullback, S., \& Leibler, A. 1951, Annals Math. Stat., 22, 79

\bibitem{Lawrence91} Lawrence, A. 1991, MNRAS, 252, 586

\bibitem{Mason06} Mason, R. E., et al. 2006, ApJ, 640, 612

\bibitem{Meisenheimer07} Meisenheimer, K., et al. 2007, A\&A, 471, 453

\bibitem{Mor09} Mor, R., Netzer, H., \& Elitzur, M. 2009, ApJ, 705, 298

\bibitem{Nenkova08a} Nenkova, M., Sirocky, M. M., Ivezi\'{c}, $\check{Z}$., \& Elitzur, M.
2008, Ap

\bibitem{Nenkova08b} Nenkova, M., Sirocky, M. M., Nikutta, R., Ivezi\'{c}, $\check{Z}$., \& Elitzur, M.
2008, ApJ, 685, 160

\bibitem{Nenkova02} Nenkova, M., Ivezi\'{c}, $\check{Z}$., \& Elitzur, M., 2002, ApJ, 570, 9

\bibitem{Nikutta09} Nikutta, R., Elitzur, M., \& Lacy, M. 2009, ApJ, 707, 1550

\bibitem{Packham05} Packham, C., Radomski, J. T., Roche, P. F., Aitken, D. K., Perlman, E., 
Alonso-Herrero, A., Colina, L., \& Telesco, C. M. 2005, ApJ, 618, L17

\bibitem{Pier93} Pier, E. A., \& Krolik, J. H. 1993, ApJ, 418, 673

\bibitem{Pier92} Pier, E. A., \& Krolik, J. H. 1992, ApJ, 401, 99

\bibitem{Raban09} Raban, D., Jaffe, W., R\"{o}ttgering, H., Meisenheimer, K., \& Tristram, K. 2009, MNRAS, 394, 1325

\bibitem{Radomski08} Radomski, J. T., et al. 2008, ApJ, 681, 141

\bibitem{Ramos11} Ramos Almeida, C., et al. 2011a, ApJ, 731, 92

\bibitem{Ramos11b} Ramos Almeida, C., et al. 2011b, MNRAS, 417, L46

\bibitem{Ramos09a} Ramos Almeida, C., et al. 2009, ApJ, 702, 1127

\bibitem{Schartmann08} Schartmann, M., et al. 2008, A\&A, 482, 67

\bibitem{Thompson09} Thompson, G. D., Levenson, N. A., Uddin, S. A., \& Sirocky, M. M. 2009, ApJ, 697, 182

\bibitem{Tristram07} Tristram, K. R. W, et al. 2007, A\&A, 474, 837

\end{thebibliography}
\end{document}